# Multi-slot optical Yagi–Uda antenna for efficient unidirectional radiation to free space

Jineun Kim[1]†, Young-Geun Roh[1]†, Sangmo Cheon[1]†, Jong-Ho Choe[2], Jongcheon Lee[1], Jaesoong Lee[1], Un Jeong Kim[1], Yeonsang Park[1]*, In Yong Song[1], Q-Han Park[2], Sung Woo Hwang[1], Kinam Kim[1] & Chang-Won Lee[1]*

[1]*Samsung Advanced Institute of Technology, Yongin-si, Gyeonggi-do 446-712, Republic of Korea*

[2]*Department of Physics, Korea University, Seoul 136-701, Republic of Korea*

†These authors contributed equally to this work.

Correspondence and requests for materials should be addressed to Y.P. (email: yeonsang.park@samsung.com) and C.-W.L (email: chang-won.lee@samsung.com).

**Abstract**

Plasmonic nanoantennas are key elements in nanophotonics capable of directing radiation or enhancing the transition rate of a quantum emitter. Slot-type magnetic-dipole nanoantennas, which are complementary structures of typical electric-dipole-type antennas, have received little attention, leaving their antenna properties largely unexplored. Here we present a novel magnetic-dipole-fed multi-slot optical Yagi–Uda antenna. By engineering the relative phase of the interacting surface plasmon polaritons between the slot elements, we demonstrate that the optical antenna exhibits highly unidirectional radiation to free space. The unique features of the slot-based magnetic nanoantenna provide a new possibility of achieving integrated features such as energy transfer from one waveguide to another by working as a future "optical via."

An optical antenna, a miniaturized analogy of a radio-frequency (RF) antenna working in the optical regime, provides a new method of manipulating visible and infrared radiation at the nanoscale owing to the strong light-matter interaction[1–4]. Various antenna functions across the optical spectrum have been realized, such as transition rate enhancement of a quantum emitter[5], enhanced photoresponse[6], sensing of weak molecular signals[7], nano-optical tweezers[8], and beam forming[9]. In addition to the enhanced field effect, control of the radiation direction in the optical spectrum has also been successfully demonstrated by submicroscale Yagi–Uda antennas[10,11]. Optical Yagi–Uda antennas use a metal nanorod that acts as an electric dipole feed with additional nanorods for the reflector or directors[12,13]. An oscillating electric current of the feed element can be realized by the resonance-induced surface plasmon polariton (SPP) current along the length of the feed nanorod.

Aperture-type optical antennas, which can be considered as a Babinet-complementary counterpart of the rod-type optical antenna, have received little attention so far[14,15]. In the RF regime, a number of slot or leaky-wave antennas connected to waveguide structures have been studied for directional radiation control[16]. However, direct miniaturization of structures optimized for RF spectra is not possible for realizing efficient optical antenna operation because metals do not show perfect electrical conductor behaviour over the optical spectrum.

A number of issues should be resolved to obtain efficient optical Yagi–Uda antennas based on aperture slots. Unlike rod-type antennas in which the oscillating electric dipole works as a feed, slot-type antennas require a resonance condition that can determine whether the feed is working as an electric dipole, magnetic dipole, or a mixed dipole[4]. Near-field interactions between rods are replaced by SPP exchange or quasi-cylindrical wave coupling between the slot elements in a slot-type antenna[17]. The geometry of an antenna based on multiple slots should be optimized by considering the intrinsic damping effect of SPPs, the substrate effect, and fabrication errors. In this article, we address all of these issues and demonstrate a novel aperture-type optical Yagi-Uda antenna that exhibits highly unidirectional radiation to free space.

Our work is different from previously studied structures for beam forming that were based on periodic nanoholes and slits to provide the interference effect[20–22]. Our antenna structure uses the resonance of the feed element, which allows incidence-angle-independent unidirectional radiation[23]. Furthermore, to discriminate pure antenna function from the radiation trapping effect due to the high refractive index substrate, we use our antenna as a free-space directional radiator. Our proposed structure represents the first optical antenna with a magnetic-dipole feed for free-space unidirectionality, which can potentially be embedded in an integrated plasmonic structure as a radiation router (optical via) or a chromatic discriminator.

**Results**

**Far-field radiation pattern.** Theoretical analysis of aperture antenna operation is often based on Love's equivalence principle or the field equivalence principle (FEP) that enables a fast search of analytical solutions including the resonance conditions[15]. Application of FEP to a narrow aperture slot made of a perfect electric conductor (PEC) on an infinite ground plane gives an analytical solution for the radiation due to the oscillating magnetic current of the aperture. In practice, however, analytic solutions are elusive due to the strong dispersive and dissipative nature of metals with a finite conductivity. Therefore, it is important to know the leading order characteristics of the nature of the slot in terms of the radiation generator[4].

Figure 1a shows the near- and far-field radiation patterns of a magnetic dipole and an electric dipole. If the slot induces a magnetic dipole at resonance, the near-field radiation should have an electric field component that is perpendicular to the length of the slot. If, instead, an electric dipole is induced across the narrow slot width, then the electric field component of the near-field radiation is parallel to the length of the slot. As a result, the observed far-field patterns of the magnetic and electric dipoles are mutually orthogonal, which makes it easy to find the leading order of the induced dipole.

The coordinate system for a rectangular metal slot is sketched in Fig. 1b with information about the θ and ϕ distributions of the radiation of a fabricated slot with a length of $l$ = 145 nm and a width of $w$ = 50 nm on a metal film with a thickness of $t$ = 180 nm. The radiation pattern of a single slot is imaged at the back focal plane for a Fourier transformed image in transmission geometry after a proper apodization transformation and polarization filtering (see Methods). Note that the transmission intensity on resonance follows $I(\theta) = I_0 \cos^2(\theta)$ (Supplementary Fig. S1). Figure 1c shows the measured Fourier-plane image of a single slot as captured by an objective lens. The measured far-field electric field component is perpendicular to the length of the slot and has a near-uniform distribution along the *x*-axis. This confirms the slot radiates as a magnetic dipole due to the resonantly excited magnetic current, regardless of the polarization and incidence angle of the irradiating laser (see Supplementary Note 1 and Fig. S2). Note that when the polarization of the incident beam is not perfectly perpendicular to the output polarizer, a bright spot is formed at the centre of the Fourier image due to the direct transmission.

**Building a Yagi-Uda slot antenna.** Even though direct application of Babinet's principle to the optimized rod-type optical Yagi–Uda antenna does not provide an optimal design for slot-type antennas, several clues for obtaining an optimal design are provided. The operating principle of the Yagi–Uda antenna is based on the fact that each element behaves like a dipolar object and that the interactions between these elements determine the overall antenna performance. This means that the roles of the local feed and auxiliary elements in a Yagi–Uda antenna do not change as long as the slots behave as dipoles. However, a question arises as to what the operating principle is if the slots interact strongly. That is, is the operating principle of the Yagi–Uda antenna still valid with respect to the assumption that each slot represents a single elemental object. For extraordinary transmission, it has been known that the interactions between slots occur through SPP exchange or quasi-cylindrical waves. However, it is not clear if an optical antenna based on multiple slots can be built following the conventional antenna operation principle. We address this issue by analyzing the relative electromagnetic phases in the slots and show that a phase analysis can be applied to a wide range of geometrical variations of our multi-slot optical antennas. Practical limitations due to imperfect fabrication techniques can be compensated for during the simulation stage, and we show that systematic finite-difference time-domain (FDTD) simulations indeed compensate for the differences between the designed and fabricated structures.

Our approach for determining the best performing optical multi-slot antenna is as follows. First, we start by searching for the resonance condition of a narrow slot aperture. Because fabricated slots less than 100 nm in width do not have a rectangular cross section, we design a more realistic modified slot structure, as shown in Fig. 2a. Second, based on the resonance condition of the slot as a feed element, we add a reflector slot to induce unidirectional radiation. We rotate the reflector slot by 45° and illuminate the antenna with a laser beam that is linearly polarized in a direction parallel to the reflector slot so that only the feed slot is excited[10]. Due to the angular difference between the feed and reflector, there are two possible configurations, which we discuss here, and we show that optimizing the feed radiation is more important than optimizing passive elements. Third, director slots are added to improve the unidirectional properties. For each process, we compare the performance of the fabricated and simulated structures.

**Resonance of a narrow aperture slot.** A slot aperture made of a PEC has a dipolar resonance wavelength $\lambda$ that is approximately two times the slot length $l$. However, a slot made of a real metal (such as Ag) illuminated by a visible wavelength shows the following modified linear relationship between the resonant wavelength and the slot length due to the finite skin depth[24]:

$$\lambda = a\,l + b\,. \tag{1}$$

Due to fabrication limitations, two additional effects need be considered: the effects of the substrate and the cross-sectional shape, which depend strongly on the fabrication method. The inset of Fig. 2a shows a cross-sectional scanning electron microscope (SEM) image of a typical single slot used in our experiment. The SEM image shows that the fabricated slot has an isosceles trapezoid shape with an additional trench in the substrate. The width of the upper aperture at the air interface is 70 nm, while the width of the bottom aperture at the quartz substrate is 30 nm.

To determine an accurate relationship between the resonant wavelength and slot length, we measured the transmission spectra of single slots with lengths of 90–170 nm (Fig. 2b). In this experiment, the width in the middle of the slot and the thickness of the silver film were fixed at 50 nm and 180 nm, respectively. By plotting the resonance wavelength against the slot length (Fig. 2c), we obtained the following linear relation between $\lambda$ and $l$:

$$\lambda = 2.4\, l + 315 \text{ (nm)}, \quad (2)$$

which has a similar form to the resonance condition of a rod-type antenna[25].

In the three-dimensional (3D) FDTD calculations to simulate the actual slot shape, we used a simplified slot geometry with a trapezoidal-shaped cross section modelled using three additional parameters, the bottom side width ($w_B$), top width ($w_T$), and depth of the trench into the substrate ($\delta$), as shown in Fig. 2a. We first changed $w_B$ while keeping $\delta$ at zero, so that the constraint $w_B = 100 - w_T$ holds. In this case, the resonance peaks of the spectra are red shifted, and the slope of Eq. (2) increases as $w_B$ decreases. This result arises from the plasmonic effect in a rectangular slot[26,27]. We then considered the effect of over-etching the substrate, i.e. a nonzero $\delta$. The resonance of the slot with fixed $w_B$ shows a blue shift, and the slope of Eq. (1) decreased as $\delta$ increased. This result arises from the effective dielectric constant of the trenched substrate decreasing as $\delta$ increases[28].

By combining these two results, we found that the transmission spectra were best fitted when $w_T = 70$ nm, $w_B = 20$ nm, and $\delta = 40$ nm:

$$\lambda_{FDTD} = 2.4\, l_{FDTD} + 305 \text{ (nm)}, \quad (3)$$

as shown in Fig. 2d. Note that the $y$-intercept of the slope in the FDTD simulation exceeds the experimental intercept by 10 nm. This arises from the round-edge effect of a real slot, which makes the effective length of the slot larger than the simulated one.

**Two-slot nanoantenna.** The design of an efficient multi-slot antenna is based on understanding the nature of the interaction between two slots when one slot acts as a feed slot to drive the other.

We chose a 145 nm feed length to study the nanoantenna function at 664 nm (red) according to the linear relation in Eq. (3). There are two main factors to consider in the geometric configurations of two slots. First, the coupling between slots leads to energy exchange through SPPs or the evanescent coupling of electromagnetic energy[12,17]. Since both SPPs and diffracted electromagnetic energy through the slots mostly propagates perpendicular to the length of the slot[29], it is readily expected that the coupling between the two slots is the strongest when they are facing each other, as shown in Fig. 3a[30].

In this configuration, however, both the feed and auxiliary slots are excited by polarized illumination. Therefore, we use a polarization selection trick by rotating the feed element by 45° with respect to the other slot and the incident polarization[10]. Since a parallel polarized beam does not excite the slot structure, the polarization of the incident laser beam is always parallel to the auxiliary slot, leading to excitation of the feed slot only. This configuration is very effective for realizing local feeding of the Yagi–Uda structure without using a local quantum source.

Tilting the feed slot in the Yagi–Uda structure gives rise to another degree of freedom for designing an optimal antenna structure. For the two-slot case, there are two possible configurations. Let us assume a reflector slot is located to the left of the feed slot. When the reflector is tilted with respect to the feed slot as in Fig. 3b, we refer to this structure as "type A". When the feed is tilted with respect to the reflector as shown in Fig. 3c, we refer to this as "type B". Type A is the case in which the reflector (or director in other multi-slot cases) is located at the surface normal of the feed. Type B is the case in which the feed is located at the surface normal of the reflector (or director). Physically, type A is optimized for maximal transfer of the SPP energy *from the feed to the reflector/director*. In contrast, type B is optimized for *maximal* *reflection of the reflector* and *maximal directing power of the director*. We note that the incident laser polarization is always parallel to the reflector slot in both type A and B antennas.

In addition to the directions of the two elements, the length of the reflector element (L) and the distance between the two elements (D) need to be decided. We performed full 3D FDTD simulations[18] for a range of D and L values: we varied D from 130 to 180 nm and L from 55 to 195 nm (see Supplementary Figs. S3-S9). These values were determined from our fabrication ranges. Figure 4a shows a 3D view of the far-field radiation pattern obtained from a near-to-far-field transformation of the two-slot antenna radiation. By choosing an appropriate set of D and L, efficient unidirectional radiation to free-space can be obtained. For a more comprehensive overview, we present a contour map of the front-to-back (FB) ratios as a function of D and L in Fig. 4b. Large FB ratios are obtained when L > 120 nm, and the FB ratio increases with decreasing D in the same region. The maximum FB ratio for the type-A antenna, 7.6 (8.8dB), is expected with L = 155 nm and D = 130 nm, which provides a condition for the best reflector operation of the second slot. We note that, unlike rod-type Yagi–Uda antennas, the optimal condition for D = 130 nm is smaller than one quarter of the wavelength of the free-space electromagnetic wave (664/4 nm = 166 nm) or the wavelength of the SPP ($\lambda_{spp}/4 = \text{Re}[\sqrt{(\varepsilon_0 + \varepsilon_1)/(\varepsilon_0 \varepsilon_1)}]\lambda_0/4 = 0.97\lambda_0/4 = 161$ nm, where $\varepsilon_0 = 1.0$ for air and $\varepsilon_1 = -17.7 + i1.13$ for silver at $\lambda_0 = 664$ nm). The optimal distance between the feed and the reflector of a rod-type Yagi–Uda structure is not simply given by the path difference between the feed and the reflector[30]. Likewise, a simple argument for optimal distance for slot-type antennas fails because of additional phases due to strongly coupled SPP waves and finite-thickness effect of the slot aperture.

Similar FDTD simulations were performed for type-B antennas (see Supplementary Figs. S3-S9). The maximum FB ratio of 6.1 (7.8dB) was obtained when L = 165 nm and D = 130 nm. Even though the optimal D and L values do not change much for both antenna types, the maximum FB ratio of the type-B antenna is slightly smaller than that of type-A antenna. As inferred from the geometrical analysis above, we find that the type-A configuration provides a better antenna performance owing to optimal SPP energy transfer from the feed to the auxiliary elements.

**Phase analysis.** To gain a more comprehensive physical insight into the aperture-type antenna design, we investigated the phase relationship between the elements. Optimal conditions for rod-type RF and optical Yagi–Uda antenna designs have revealed that the spacing between the feed and the reflector should be one quarter of the wavelength. The spacing between directors is a more complicated parameter and is known to depend on the number and lengths of the elements[16]. It is evident that direct conversion to a Babinet-complementary geometry does not necessarily provide an optimal design for an aperture antenna. However, we show that the physical principle of the Yagi–Uda antenna still holds. In a narrow rectangular slot, the magnitude of the y-component of the electric field ($E_y$) is relatively small, and therefore the effective magnetic current density ($M_{eff}$) is generated mostly from the x-component of the electric field ($E_x$) given by $\vec{M}_{eff} = -2\hat{n} \times \vec{E} = -2\hat{z} \times \vec{E} = -2(\hat{y}E_x - \hat{x}E_y) \approx -2\hat{y}E_x$[15,31]. FDTD simulations show that the y-component of the electric field can indeed be neglected because its magnitude is almost 10 times smaller than that of the x-component (Supplementary Note 2). The phase of the field inside the slot is also uniform, and therefore, the phase of the x-component of the electric field is a leading order candidate for characterizing the magnetic currents corresponding to each slot.

We calculated the phase values at the centre of the type-A slots and plotted the difference as a function of D and L, as shown in Fig. 4c. Here, the phase difference is defined as the phase of the second slot element minus that of the feed. When the phase difference approaches −90°, electromagnetic waves radiating from the feed interfere with those reflected from the second slot to the left, resulting in unidirectional radiation towards the right-hand side of the feed. If we compare Fig. 4b and 4c, the maximum FB ratio region coincides with the region of the −90° phase difference, which is denoted by the black lines (See Supplementary Fig. S10 for type B). This results show that the phase analysis used in RF Yagi–Uda antenna theory still holds when the phase of the slot is properly defined as an object characterizing each individual slot.

We examined the two-slot devices more closely for L values of 55–195 nm while keeping D fixed at 130 nm. Figure 4d shows the azimuthal angle ($\phi_{max}$) at the maximum intensity taken from the measured Fourier-plane images as a function of L. The black and red rectangles indicate the measured and simulated $\phi_{max}$ values, respectively. Both the measurement and the simulation show a jump from $\phi_{max}$ = 360° to $\phi_{max}$ = 180° near L = 95 nm, suggesting that the mode of the two-slot configuration changes from the reflector to the director; this in turn suggests that the two-slot combination can also be considered as a feed-director. The inset of Fig. 4d shows a representative Fourier-plane image measured from the device with L = 155 nm and D = 130 nm (reflector condition). These measured data experimentally demonstrate that the phase matching condition adapted from the RF antenna can also be applied to find the optimal reflector of the optical nanoantenna. The director condition corresponding to the phase difference of −270°, however, could not be found within the given ranges of D and L. The best director condition we found in Fig. 4 was D = 130 nm and L = 95 nm, with the largest phase difference of −170° as denoted by the white lines in Fig. 4b and c. Therefore, a single director is not sufficient for directional radiation, which in turn, implies multiple slots will show better antenna performance.

**Nanoantennas with more than two slots.** To show the enhancement in the directivity with more directors, three- and five-element slot antennas were designed and fabricated (Supplementary Notes 3 and 4; Supplementary Figs. S11-S16). To examine the fundamental antenna behaviour independent of the fabrication errors, more than 7 antenna structures with the same geometry were prepared and measured. For the three-slot antenna, we found that a structure with a reflector of $L_{refl}$ = 155 nm and $D_{refl}$ = 130 nm and a director of $L_{dir}$ = 95 nm and $D_{dir}$ = 140 nm provided the largest FB ratio for a feed of $L_{feed}$ = 145 nm (see Supplementary Note 3 and Fig. S12). For the five-slot antenna, we further optimized the geometrical factors of an additional two director slots based on the three-slot antenna structure. We note that the five-slot antenna with a feed, reflector, and three directors becomes a complementary structure of the previously reported Yagi–Uda nanoantenna[10,11]. Similar to the two-slot case, multiple slots with reflector or director elements tilted by 45° with respect to the feed allow for two possible configurations. For type-A antennas, the reflector and director slots are tilted with respect to the feed maximizing the interaction of SPPs generated by the feed, as shown in the inset of Fig. 5a. The measured FB ratios from individual type-A antenna structures are plotted in Fig. 5a as a function of the number of slots. In general, multiple slots show better unidirectional performance on average, even though variations from individual structures become large. The best FB ratio from a type-A antenna was 8.5 (9.2dB).

We also prepared and measured type-B structures, which are complimentary structures of those previously reported[10]. Unlike type A, only the feed slot is tilted while all the other elements remain orthogonal to the entire antenna structure, as shown in the inset of Fig. 5b. In this case, the direction of the radiation from the feed is not aligned in the direction of the auxiliary elements. However, the interaction between the feed and other elements is enough to generate unidirectional radiation. A similar optimization procedure was performed to find the D and L values for the maximum FB ratio (Supplementary Note 3 and Fig. S13). The measured FB ratios obtained from the type-B slot antennas are plotted in Fig. 5b as a function of the number of slots. The maximum FB ratio of 2.6 (4.1dB) was obtained when $L_{refl}$ = 165 nm and $D_{refl}$ = 130 nm for the reflector and $L_{dir}$ = 95 nm and $D_{dir}$ = 150 nm for the director. The average FB ratios of the type-B antennas are, however, ~50% smaller than those of the type-A antennas. Even though the FB ratios of the type-B antennas might increase slightly with different choices of D and L, we find that the type-A antennas generally produced relatively larger FB ratios.

For both antenna types, the FB ratio and deviation increases as the number of slots increases. (Supplementary Note 4; Supplementary Figs. S15-S17)With an increasing number of slots, the radiation pattern also becomes narrower with an increasing FB ratio. Figure 5c shows the Fourier-plane image obtained from one of the type-A antennas with the maximum FB ratio, and Fig. 5d shows the corresponding simulation result. The radiation pattern is tilted slightly to the lower right-hand corner away from the axis passing through the centres of all the slots. The half power beamwidth (HPBW), which is defined by the angle between the half-power points of the main lobe in a plane, could not be directly measured due to the limited numerical aperture of the objective in the θ coordinate. A HPBW value of 71°, however, could be obtained from the simulated result. We note that the HPBW should be evaluated in the plane containing the maximum lobe intensity at $(\theta_{max}, \phi_{max})$ = (56°, 347°), and we took the two −3-dB points from the same ϕ = 347° plane for the HPBW extraction.

**Outlook**

In conclusion, we have demonstrated the first magnetic dipole feed-driven multi-slot nanoantenna for unidirectional free-space radiation in the visible wavelength range. We achieved a maximum FB ratio of 8.5 (9.2 dB) for a five-slot antenna. Even with the limited fabrication margins and a lack of methods for performing a full optimization, we were able to show that a leading-order phase analysis can still be applied to the design of efficient aperture-type optical Yagi–Uda antennas. Better unidirectional radiation performance can be anticipated with further optimization of the antenna structures and materials.

Our optical antenna structure delivers a few structural advantages for integrated nanophotonic structures. First, SPPs can be directly injected to the antenna either by a thin metal layer or a metal-insulator-metal waveguide, as illustrated in Fig. 5(e). Second, spatial separation owing to the metallic layer between the upper and lower hemispheres allows a 3D optical via between physically separated layers. Third, the metal layer can be simultaneously used as electrodes for electrically excitable integrated optical devices. Other imminent subsequent works, based on this multi-slot nanoantenna structure, are a beam router driven by the polarization of the incident beam and slot antennas driven by local sources such as quantum dots or dye molecules. Elastic and inelastic single molecule spectroscopy using multi-slots is expected to provide new types of plasmonic enhancement that could lead to various chemistry and biology related research.

**Methods**

**Sample preparation.** A 180-nm-thick Ag film was coated by electron beam evaporation onto a 250-μm-thick quartz substrate, with an additional 10-nm-thick $MgF_2$ coating to prevent oxidization. The slots were formed by focused ion beam (FIB) milling using the FEI Helios NanoLab. Each fabricated nanoantenna structure was at a distance of at least 50 μm from the others to prevent inter-device interactions. The feed element's shape, defined by the film thickness, length, and width, was designed by considering the optical separation of the Fabry–Pérot mode due to two interfaces—air-metal and metal-substrate.

**Optical measurement.** We imaged the radiation from a narrow single slot in the back focal plane (i.e. Fourier or *k*-space imaging) in a microscope setup. For the transmission spectra or Fourier plane imaging measurement, a 664-nm polarized laser diode with a Glan–Taylor polarizer was used to illuminate the slot from the substrate side. For the magnetic dipole radiation of a single-slot experiment, several polarizations of the incident beam were tested to confirm the resonant feature of the slot radiation. For multi-slot antenna experiments, the incident polarization was set always parallel to the reflector or director slots to excite the feed slot only. The incident beam is focused by a 20× objective lens to form a Gaussian beam on the sample. The transmitted beam is collected by a 100× objective lens with a numerical aperture (NA) of 0.95. The apodization of the microscope objective is calibrated by varying the off-axis position of the laser beam[32]. Another Glan–Taylor polarizer located in the (real space) image plane (analyzer) is set after the microscope. For the single-slot experiment, the analyzer polarization direction is perpendicular to the length of the slot. For multi-slot experiments, the polarization of the transmitted beam is perpendicular to the incident laser beam and parallel to the reflector or director slots. This polarization and analyzer setup allows us to block direct laser transmission and let the detector measure only the radiation from the whole antenna structure.

The light collected at another port of the microscope was diffracted using an imaging spectrometer with an f-number of f/4 and was captured by an electron-multiplying charge-coupled device (EMCCD). The spectral peak position of the transmitted beam corresponding to the slot resonance does not change with angle, guaranteeing that the slot mode still forms for the polarized incident beam. For real and Fourier plane imaging, we use the left port of the microscope with another Glan–Taylor polarizer and a scientific CMOS camera. The exposure time was controlled to prevent intensity saturation and determine the FB ratio.

**Electromagnetic simulation.** We used Lumerical for the FDTD simulations to find the near-field profile of the entire antenna structure. A volume of $8 \times 8 \times 4$ μm$^3$ was used. To obtain the far-field radiation patterns from the FDTD results, we applied the Schelkunoff equivalence principle on a closed boundary including the antenna structure and translated the currents onto the far-field surface using a Green's function. Instead of using the free-space Green's function, we used reciprocity theorem and the responses of the layered system under plane wave illumination to calculate the far-field radiation[32]. We obtained the far-field information of the layered system at a distance of $r = 1$ m from the centre of our nanoantenna.

**Acknowledgements**

The authors are grateful to Prof. M. L. Brongersma at Stanford University for stimulating and fruitful discussions.

**Author contributions**

C.-W.L. initiated the project and conceived the experiments. Y.-G.R. and Y.P. fabricated the devices. J. K., Y.-G.R., and J.L.(2) performed the optical measurements. J. K., S. C. J.L.(1), and Y.P. analyzed data. S.C., J.-H.C., and Q.-H.P. provided modelling and theoretical foundation. U.J.K. and I.Y.S. participated in fabrication process. Y.P. and C.-W.L. wrote the manuscript. S.W.H. and K.K. provided suggestions throughout the project. All authors commented on the manuscript.

**Additional information**

**Supplementary Information** accompanies this paper at

www.nature.com/naturecommunications

**Competing financial interests:** The authors declare no competing financial interests.

**Reprints and permission information** is available online at

http://npg.nature.com/reprintsandpermissions/

Figure Captions

**Figure 1 | Dipole moment of a single slot and the corresponding radiation pattern.** (**a**) Near-field radiation pattern in free space from a slot when a magnetic dipole is formed (top left) and when an electric dipole is formed (top right). The far-field radiation pattern of the corresponding field is shown underneath. (**b**) The coordinate system used to measure the radiation pattern from a single slot. The Fourier-plane image of the slot (or multi-slot antenna) is recorded as the intensity distribution projected onto a hemisphere with a sufficiently large radius surrounding the entire structure emitting the light. (**c**) A representative Fourier-plane image of the radiation measured from the single slot with $l$ = 145 nm with 664-nm incident light. Two symmetric lobes are formed equally along the direction normal to the length of the slot similar to the simulation described in Supplementary Fig. S2.

**Figure 2 | Optical characteristics of a single slot.** (**a**) Schematic of the simulated slot with the shape parameters. The slot has a trapezoidal-shaped cross section with shorter bottom width ($w_B$) and a longer top width ($w_T$). The depth of the trench into the substrate is δ. (Inset) Cross-sectional SEM image of the Ag single slot fabricated by focused ion beam (FIB) milling used in our experiment. The scale bar corresponds to 100 nm. The trapezoidal-shaped cross section and trench into the substrate can be observed. (**b**) Measured transmission spectra in the visible wavelength of single slots with lengths of 90–170 nm. The incident light is polarized normal to the length of the slot. The peak position, denoted by a red arrow, represents the resonance wavelength in each spectrum. (**c**) Resonance wavelength from the measured transmission spectra of a single slot on a glass substrate with various lengths ($l$). The linearity between the slot length and peak position is described by λ = 2.4 $l$ + 315 nm and shown in red line. This relation gives a design rule for single-slot resonance and a basic guideline for designing a multi-slot antenna with directionality. (**d**) Resonant wavelength from the simulated transmission spectra of the trapezoidal-shaped slot on the glass substrate with various lengths for $t$ = 180 nm, $w_T$ = 70 nm, $w_B$ = 20 nm, and δ = 40 nm. A linear relation between the slot length and peak position can also be obtained: $\lambda_{FDTD}$ = 2.4 $l_{FDTD}$ + 305 (nm).

**Figure 3 | Two-slot configurations.** (**a**) Geometric configuration of two slots. Two parallel slots have the strongest interaction when they are located face-to-face. However, the linearly polarized incident laser can then excite both slots. Rotating one slot by 45° with respect to the other produces two possible configurations. (**b,c**) Two configurations for a two-slot antenna when one slot is tilted by 45° with respect to the length direction of the feed. Type A is shown in (**b**) and type B is shown in (**c**). The length of the second slot is denoted as L, and the distance between the two elements as D. The polarization of the incident and transmitted light are indicated by the blue arrow. Type A is a structure in which the centre of the second tilted slot is aligned with the normal direction of the feed. Type B is a structure in which the centre of the feed is aligned with the normal direction of the second tilted slot. A clockwise-rotated configuration of (**c**) is also shown for the convenience of the FDTD simulation. The dashed blue lines represent the radiation direction of feed slot.

**Figure 4 | Antenna characteristics of two-slot devices.** (**a**) Simulated far-field pattern from a two-slot antenna. (**b**) FB ratio contour plot as a function of L and D of the reflector for a given feed with a 145 nm length. The maximum FB ratio exists around the region of L = 155 nm and D = 130 nm. The black (white) line denotes a phase difference of −90° (−170°) between the feed and the additional slot obtained from the phase map in (**c**). (**c**) The phase difference between two slots. The phase difference of −90° is indicated by the black line, where the second slot acts as a reflector (backward light from the slots is out of phase). (**d**) The $\phi$ angles with maximum intensity ($\phi_{max}$) as a function of L. For $75 \leq L \leq 95$ nm, $\phi_{max}$ changes from 150° to 330°. This plot shows that the tilted second slot works as a reflector when L > 100 nm and as a "weak" director when L < 100 nm. The black (red) rectangle indicates the $\phi_{max}$ obtained from the experiment (simulation). The inset in (**d**) shows a typical Fourier-plane image of the radiation of two-slot antenna for L = 155 nm and D = 130 nm. The antenna radiates light directed to $\phi_{max} \sim 330°$, i.e. forward, which shows clearly that the second slot acts as a good reflector.

**Figure 5 | Performance of multi-slot nanoantenna configured as type A and type B.** (**a**) Measured FB ratios of the type-A antennas as a function of the number of slots. (**b**) Measured FB ratios of the type-B antennas (complementary structure of the previously reported Yagi–Uda antenna). SEM images of each type of antenna are shown in the insets (scale bars: 200 nm). The black symbols denote the FB ratios measured from 28 devices (7 devices for each antenna configuration). The red rectangles denote the average FB ratio of each case. As the number of slots increases, the average FB ratios and the deviation increases. The maximum FB ratio obtained is 8.5 (9.2 dB) from a five-slot antenna. The average FB ratio of the type-B antenna is ~50% smaller than that of the type-A antenna. (**c**) The measured Fourier-plane image (colour plot), and the radiation patterns in θ and $\phi$. The image is taken from the best type-A antenna with five elements. (**d**) The simulated antenna radiation pattern of the antenna with the same five-slot geometry. (**e**) Schematic view of a multi-slot nanoantenna with a plasmonic waveguide.

Figure 1.

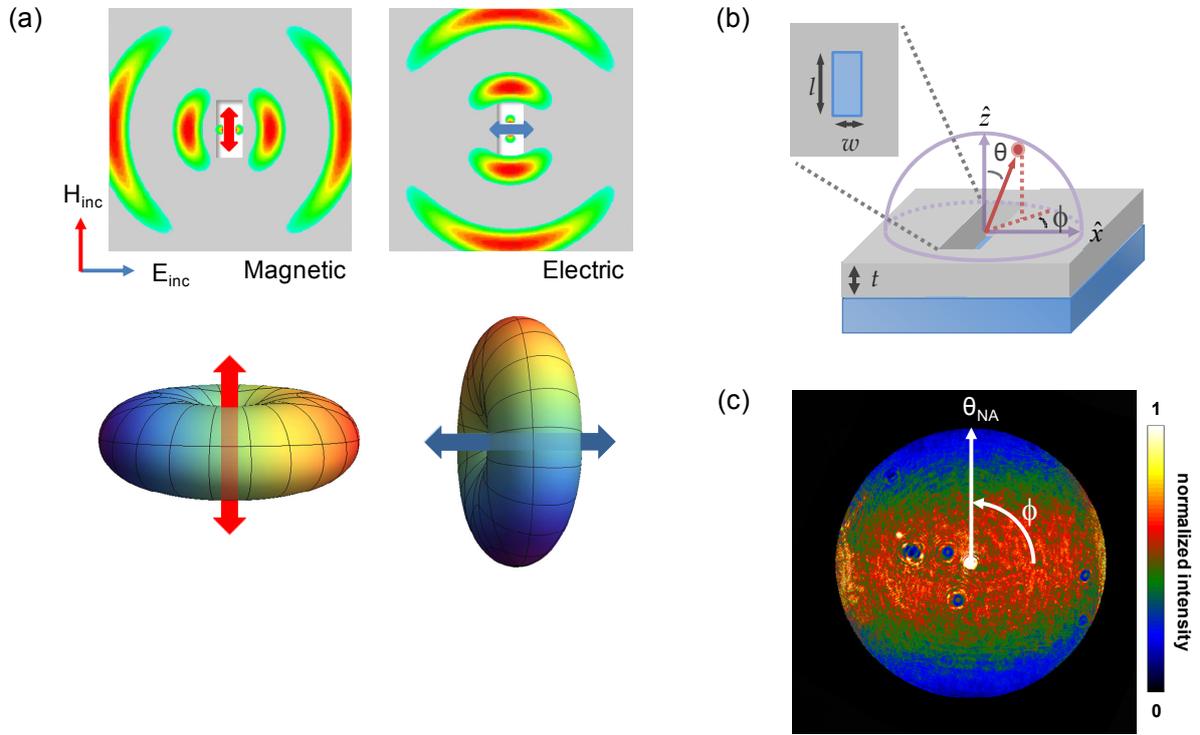

Figure 2.

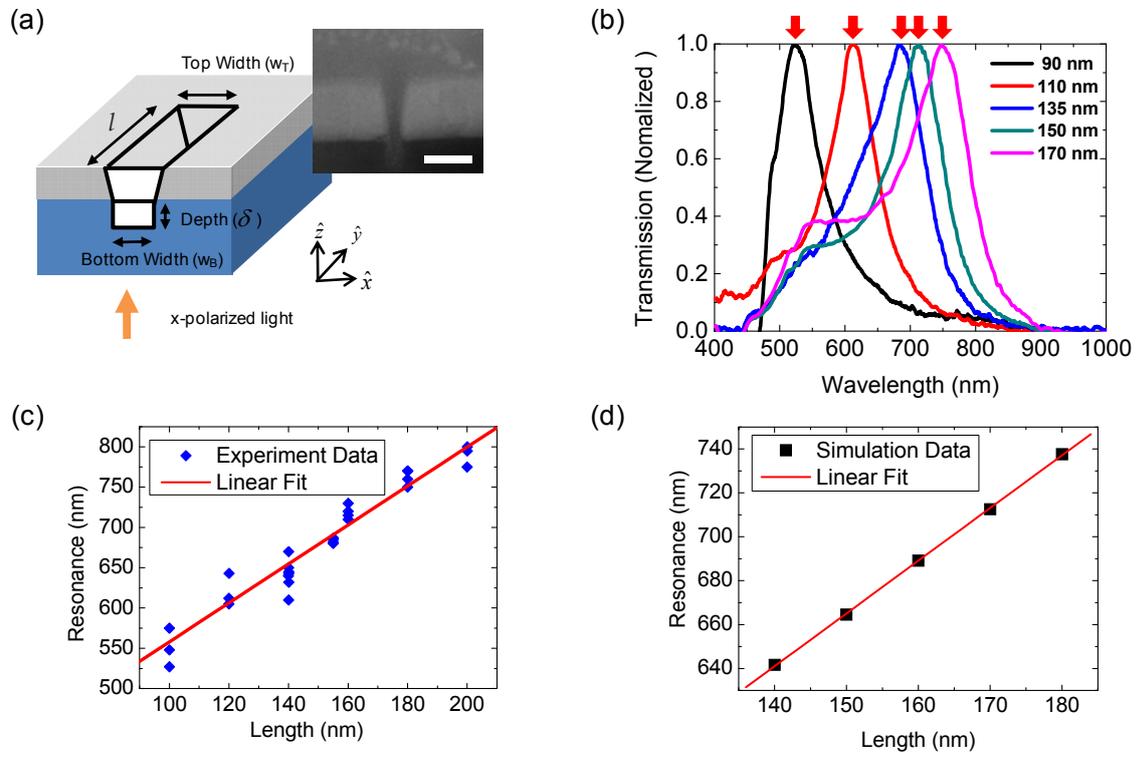

Figure 3.

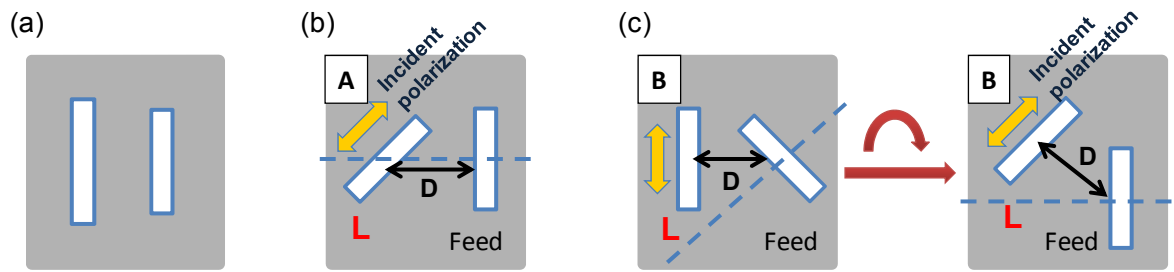

Figure 4.

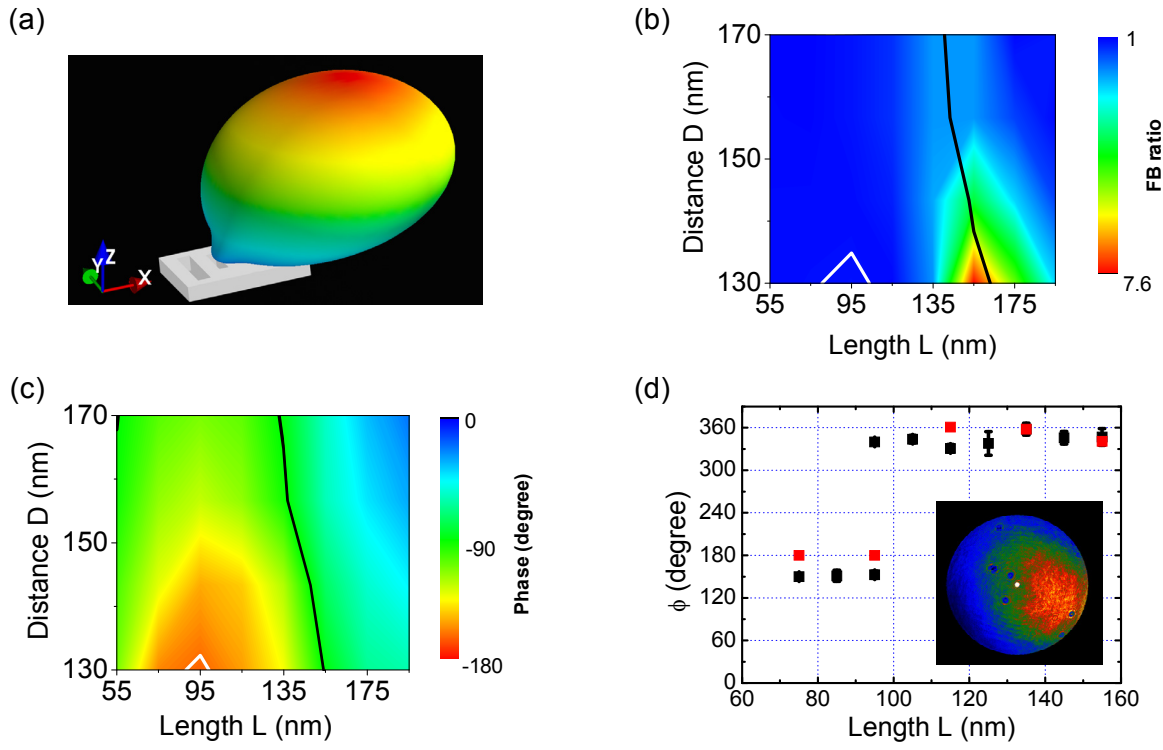

Figure 5.

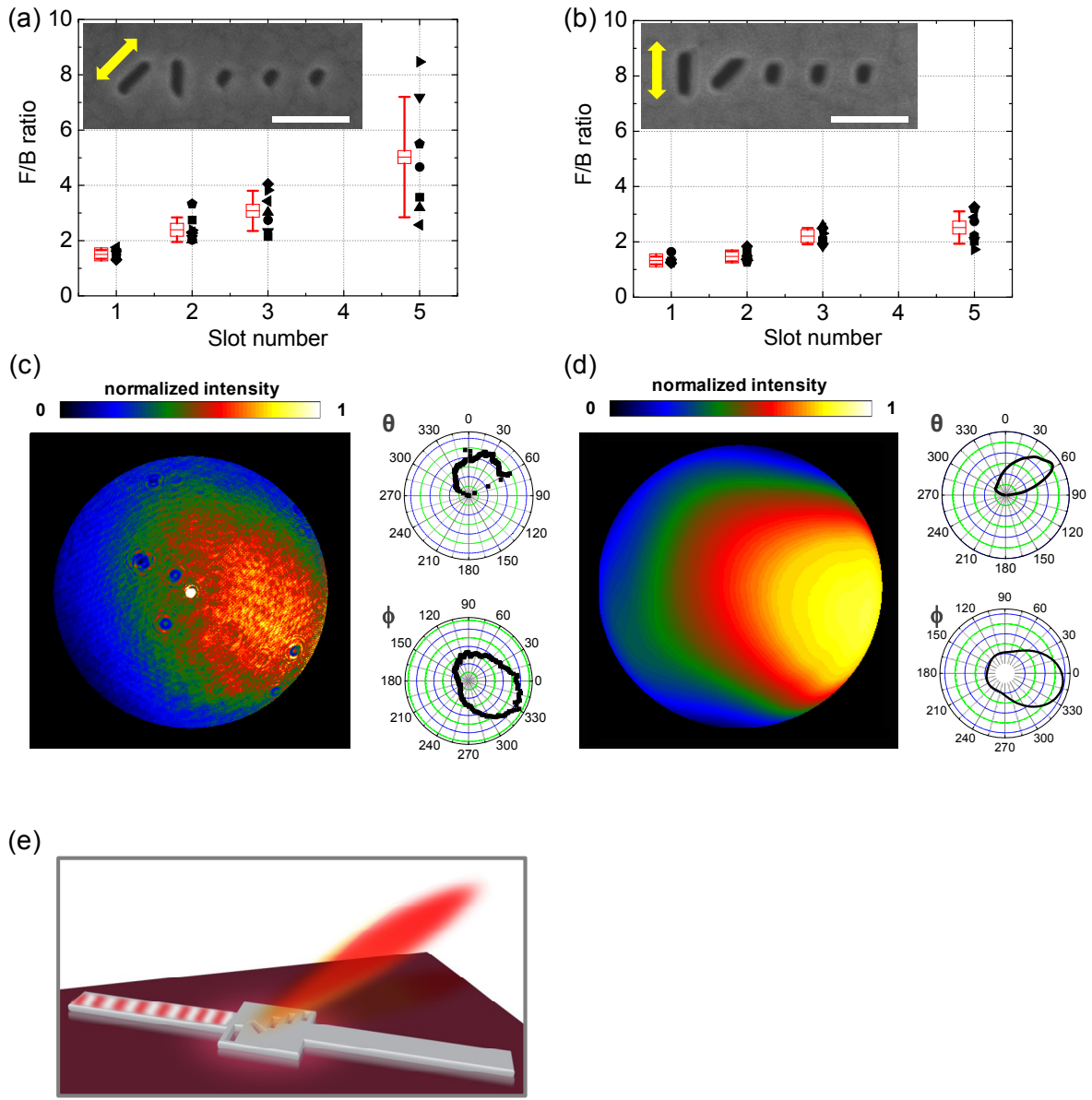

# Supplementary Information

# Multi-slot optical Yagi–Uda antenna for efficient unidirectional radiation to free space

Jineun Kim[1]†, Young-Geun Roh[1]†, Sangmo Cheon[1]†, Jong-Ho Choe[2], Jongcheon Lee[1], Jaesoong Lee[1], Un Jeong Kim[1], Yeonsang Park[1]*, In Yong Song[1], Q-Han Park[2], Sung Woo Hwang[1], Kinam Kim[1] & Chang-Won Lee[1]*

[1]*Samsung Advanced Institute of Technology, Yongin-si, Gyeonggi-do 446-712, Republic of Korea*

[2]*Department of Physics, Korea University, Seoul 136-701, Republic of Korea*

†These authors contributed equally to this work.
Correspondence and requests for materials should be addressed to Y.P. (email: yeonsang.park@samsung.com) and C.-W.L (email: chang-won.lee@samsung.com).

**Supplementary Notes**

**1. Single slot: polarization dependence and equivalence between a magnetic dipole and a single slot**

It is well known that no mode is formed by an incident plane wave with parallel polarization along a PEC rectangular slot. In our multi-element slot antenna, the feed element is rotated by 45° with respect to the polarization of the incident plane wave, which only excites the feed slot. Other antenna elements are aligned parallel to the polarization of the incident plane wave to minimize direct transmission. To verify the polarization-dependent mode excitation at optical frequencies, polarization-dependent transmission data were obtained through FDTD simulations and experiments. The notation for the polarized wave is shown in Supplementary Fig. S1a. As shown in Supplementary Fig. S1b, the transmission intensity at the resonance wavelength has a cosine-square dependence on the polarization of the incident plane wave; this result is expected from the cosine projection of an electric field in radio-frequency (RF) antenna theory and extraordinary transmission of a single subwavelength slot.

From Babinet's principle, a PEC single-slot antenna can be treated as a magnetic dipole in leading order[1]. Furthermore, a recent study shows that Babinet's principle can be applied to metamaterial design even with non-PEC materials[2]. To verify the equivalence of a magnetic dipole and our Ag slot even when the slot is excited by a 45° polarized plane wave, we compared the radiation patterns and field components of (a) an ideal magnetic dipole, (b) a PEC rectangular slot, (c) a 155-nm-long Ag slot under a perpendicularly polarized plane wave, and (d) the same Ag slot excited by a plane wave with 45° rotated polarization. Here, the magnetic dipole source oscillates with a wavelength corresponding to 664 nm, and the PEC and Ag slots have a resonance mode of 664 nm.

To obtain far-field radiation patterns from the FDTD results, we apply the Schelkunoff equivalence principle on a closed boundary including the antenna structure and translate the currents on the surface to the far-field using Green's function[3]. Usually, $e^{i\vec{k}\cdot\vec{r}}/r$ is chosen as an appropriate form of the Green's function when the antenna structure is isolated and surrounded completely by a closed boundary surface. However, our system of interest is made of slot antennas on a substrate, which makes the closed box cross a number of layer interfaces. Therefore, direct application of the free-space Green's function to the currents on the surface would not give correct answers due to reflection effects at the interfaces between each layer. Therefore, instead of using the free-space Green's function, we used reciprocity theorem and the responses of the layered system under plane wave illumination to calculate the far-field radiation[4]. We obtained far-field information in the layered system at a distance of $r = 1$ m from the centre of our nanoantenna.

The calculated far-field radiation patterns are shown in Supplementary Fig. S2. Note that a near-to-far transformation is performed in the upper hemisphere, where the slot is placed at the origin, and the radiation wave propagates upward. We plot (from top left to bottom right) (1) $|\vec{E}|^2$, (2) $|\vec{E}|^2$ with an *x*-polarizer, (3) $|\vec{E}|^2$ with a *y*-polarizer, (4) $|E_x|$, (5) $|E_y|$, and (6) $|E_z|$ in the far-field regime (1 m distance from the radiators). The magnetic field components can be calculated by taking curl values of the electric field components. In the radiation patterns from a magnetic dipole, the magnetic field component must be parallel to the length of the dipole, whereas the electric field components are perpendicular to the length of the dipole, as shown in Supplementary Fig. S2. The experimental data shown in Fig. 1c in the main text show exactly the same radiation patterns as Supplementary Fig. S2 c-2 (or d-2). Therefore, we confirm that the radiation from a Ag rectangular slot excited by a 45° polarized plane wave is identical to the ideal magnetic dipole.

## 2. Two slots: interaction between plasmonic slots

In this section, we provide a comprehensive analysis of the integration of two tilted slots, a problem that has attracted little research attention thus far. We refer to the excited slot, which is rotated by 45° with respect to the other slots and the incident plane wave, as the feed slot; the other slot is referred to as the second slot. Here, analyses of the electromagnetic mode formation of each slot, radiation patterns, charge distributions, and phase distributions for various geometries of the second slot are presented as a function of the centre-to-centre distances (D) and lengths (L).

### 2-1. Definition of the front-to-back (FB) ratio

We systematically studied the two-slot configuration for various D and L values of the second slot. To quantitatively compare the experimental and simulated data and assess the antenna performance, a figure of merit is required. The most commonly used figure of merit is the front-to-back (FB) ratio. Generally, the FB ratio is given by the ratio of the power of the maximal radiation lobe (forward lobe) to the power of the lobe formed on the opposite side (back lobe)[5]. However, the back lobe cannot be clearly distinguished in some cases; in such cases, other definitions of the FB ratio may be used. For example, for given θ with respect to the forward lobe, the intensity ratio between the point of maximum emitted power in the ϕ coordinate and the point of maximum power in the opposite ϕ direction in the radiation pattern can be taken. For the measured Fourier space images of all the fabricated nanoantennas and for the simulated data, we have used this definition. To convert the measured intensity to power, area-integrated lobe intensities with an approximately 3° spherical surface in (θ, ϕ) coordinates are used.

### 2-2. Mode formation

Mode excitation inside a feed slot at resonance should occur regardless of the existence of an adjacent slot under incident polarization parallel to the adjacent slot. To verify this, we plot the electric field intensities of the slots irradiated by a polarized incident laser. As shown in Supplementary Fig. S3, a strong electromagnetic mode indeed forms inside the feed slot in all cases. When D = 130 nm and L > 135 nm, weak mode formation in the second slot can be observed suggesting that coupling emerges between the two slots. This coupled mode formation would result in strong directional radiation.

### 2-3. Directionality: far-field patterns, θ-plot, and ϕ-plot

We compare the radiation patterns in three different ways. The sample surface is in the *x–y* plane in the spherical coordinate system. We place the feed along the *y*-axis. We then draw the density plots of the polarization-filtered radiation intensities in the upper hemisphere (of air) with *r* = 1 m using the near-to-far-field transformation of the FDTD results from a rectangular parallelepiped enclosing the whole nanoantenna structure.

By varying the values of L and D, we search for conditions of optimal directional radiation of the two slots of type A and B, respectively. We plot the radiation patterns in a density plot format to find the ($\theta$, $\phi$) coordinate at the maximum lobe intensity and FB ratio, as shown in Supplementary Fig. S4. The number in each plot shows the FB ratio of the corresponding D and L combinations. For a type-A reflector, we find that L = 155 nm and D = 130 nm provides the maximum FB ratio, reaching 7.6 (8.8 dB) with only two slots. For a type-B reflector, we find that L = 165 nm and D = 130 nm provides an FB ratio of 5.8 (7.6 dB). Due to complex correlations between D and L for optimal directionality, it is necessary to search for combined conditions of D and L. We draw a $\theta$ polar plot at optimum $\phi$ (Supplementary Fig. S5) to find the radiation angle from the metal surface. We also draw a $\phi$ polar plot at optimum $\theta$ (Supplementary Fig. S6) to find the maximal directional angle, according to the reflector or director function of the adjacent slot. Surprisingly, the $\theta$ or $\phi$ values at the maximal lobe intensities are stable with comparably large changes in D and L, as shown in Supplementary Fig. S7. This stability of the directional angle shows that our design would lead to robust experimental results even with fabrication and measurement errors. However, we note that the conditions in terms of D and L for a good director cannot be found within our D and L ranges. The charge distribution and phase analysis presented in the next section showed that a good director condition could not be obtained with only a two-slot geometry.

## 2-4. Charge distribution and phase

To determine the nature of the coupling between the feed and the adjacent slot, we analyze the charge density distribution of our antenna slots, as shown in Supplementary Fig. S8. Our optimal reflector and director (red and blue square boxes) show capacitive and inductive coupling respectively, which is opposite to the couplings in a multiple-rod-based RF antenna. This is understood as a result expected from Babinet's principle, which states that the electric and magnetic fields are exchanged for a pair of complementary structures. We see that the coupling exactly follows the trend of the directional radiation obtained in Fig. S4. However, simply observing the charge configuration is not sufficient to find optimal conditions for the reflector and director. Therefore, we analyze the phase distribution of the *x*-component of the electric field, as shown in Supplementary Fig. S9. The magnitude and phase of the magnetic dipole current generated at a rectangular slot can be calculated mainly from the electric field component at the opening of the slot[5]. The effective magnetic current of a slot is given by $\vec{M}_{eff} = -2\hat{n} \times \vec{E} = -2\hat{z} \times \vec{E} = -2(\hat{y}E_x - \hat{x}E_y) \approx -2\hat{y}E_x$, since $|E_x| >> |E_y|$ in all cases. In other words, compared to the *x*-component, the *y*- and *z*-components are very weakly coupled to the feed slots[6]. Therefore, we chose the *x*-component of the electric field at the centre of each slot to define the phase. We find that the phases inside the slots are almost uniform irrespective of the tilting angle, and therefore, we can define a single-valued phase by taking the value at the centre of each slot. The uniformity of the phase inside a slot suggests that each slot can be regarded as a single magnetic dipole object, and this provides the groundwork for the method of moments in antenna theory.

Supplementary Fig. S10 shows contour plots of the calculated phase difference and FB ratio for various D and L values. We found that the phase difference, which is defined as the phase of the second slot element minus that of the feed, and the FB ratio have a close relationship. For a type-A antenna with fixed L = 155 nm and decreasing D, the phase decreases from −90°, which means that the antenna performance worsens. For fixed D = 130 nm, the phase decreases from −60° to −170° and then increases with decreasing L. Similar behaviour can be observed for type-B antennas. Note that the biggest FB ratio for the type-A antenna occurs when D = 130 nm and L = 155 nm. The biggest FB ratio for the type-B antenna occurs when D = 130 nm and L = 165 nm. We find that the phase difference for these D and L combinations of both antenna types is −90°, suggesting constructive interference of electromagnetic waves in the forward (right) direction and corresponding phase lag in the phasor description. A good director should have a phase difference of −270°. However, the best phase that we could obtain for a director was only −170° for both type-A and -B antennas within a limited range of values of D (130–170 nm) and L (55–195 nm). There are two reasons for not being able to obtain sufficient phase values for directors: (1) it is not possible to obtain a D of less than ~130 nm, which makes the two slots merge together, and (2) a smaller L leads to another mode formation of the second slot parallel to the incident light, which means that the second slot can no longer be considered as a narrow rectangular slot.

## 3. Three slots

The antenna operation of two slots composed of a feed and a reflector can be enhanced by the addition of more director slots. We consider here a three-slot antenna. As with the two-slot antenna, we studied two types of configurations: type A and type B (see Supplementary Fig. S11a and b, respectively). In the case of three slots, there are five variable parameters: the lengths of the three slot elements ($a_1$, $a_2$, and $a_3$) and the two nearest-neighbour distances between the slots ($b_1$ and $b_2$). To examine the antenna performance in terms of the FB ratio and half power beamwidth (HPBW) of the forward lobe, we considered 12 different sets of these parameters for each type using FDTD simulations. Here, the HPBW tells us how narrow the lobe is in space. We fixed $b_1$ = 130 nm for a given feed length of $a_2$ = 145 nm, which was obtained from the two-slot experiments and simulations for the best reflector-feed spacing, and we scanned over two different reflector and feed lengths ($a_1$ and $a_3$) and three different feed-director spacings ($b_2$). We found the best allowable parameters for a type-A antenna to be $a_1$ = 165 nm, $a_3$ = 95 nm, and $b_2$ = 140 nm. Similarly, we found the best allowable parameters for a type-B antenna to be $a_1$ = 165 nm, $a_3$ = 95 nm, $b_1$ = 130 nm, and $b_2$ = 150 nm. Most importantly, type-A nanoantennas almost always show better FB ratios than type B with various geometries. We plot the ratios of the FB ratio of type A to that of type B in Supplementary Fig. S12c, which shows the superior antenna performance of type A over type B.

The simulated radiation patterns of the optimal type-A antenna structure are drawn in Supplementary Fig. S12a, b, and c. Comparing these with those of the two-slot antenna, the three-slot antennas show better directionality and a reduced HPBW. Supplementary Fig. S12d shows a well-formed fundamental mode inside the feed as well as a mode in the reflector. Even though the mode in the director does not seem to form, a significant charge distribution (Supplementary Fig. S12e) and phase distribution (Supplementary Fig. S12f) builds in the director and contributes towards better directional radiation. Similarly, Supplementary Fig. S13 shows the radiation patterns, mode, charge distribution, and phase distribution for type B.

**4. Five slots**

The full optimization of a five-slot antenna requires a large number of combinations of parameters, making global optimization an extremely time-consuming task. To extract the effect of the increasing number of director slots, we chose 32 different parameter sets for simulations of type-A and type-B antennas. Starting from the three-slot parameters, we added two more directors with the same length and distance. This approach is commonly used in the parameter sweeping method in RF antenna theory[5]. As with the two- and three-slot antennas, we studied two types of configurations: type A and type B (see Supplementary Fig. S14a and S14b, respectively). The table in Supplementary Fig. S14c shows our scanned parameters for various $b_1$, $b_2$, and $b_3$ distances for $a_1$ = 155 or 165 nm and $a_3$ = 95 or 115 nm. We observe that the main forward lobe is narrower than the optimal two-slot antenna and that type-A antennas still provide larger FB ratios than type-B antennas, as shown in Supplementary Fig S14d.

Both experimentally measured and simulated data show that type A always provides higher FB ratios than type B does, as described in the main text. However, there is room for further optimization of the five-slot nanoantennas, and we intend to present this in future work. The radiation patterns in Supplementary Fig. S15a, b, and c show the radiation patterns in the density plot, $\theta$−plot, and $\phi$−plot, respectively. Supplementary Fig. S15d shows the fundamental mode inside the feed slot and the modes in other slots. Supplementary Fig. S15e and f show the charge and phase distribution maps, respectively, of a type-A antenna. Note that uniform phases are formed in the nearest-neighbour slots to the feed only and not in farther slots. Similarly, Supplementary Fig. S16 shows the same results of a type-B antenna. We also note that a type-B five-slot antenna is a complementary structure of the previously studied Yagi–Uda nanoantenna[7].

Finally, to compare the overall performance of the two-, three-, and five-slot nanoantennas, the simulated Fourier images and $\theta$ and $\phi$ plots are shown in Supplementary Fig. S17. We chose the best cases considered in the previous sections. We derived the following conclusions: (1) more directors lead to narrower radiation lobes in $\theta$ and $\phi$, as shown in Supplementary Fig. S17a and b, and (2) type A generally produces larger FB ratios than type B does for all cases.

**Supplementary References**

# Supplementary Figures

(a) 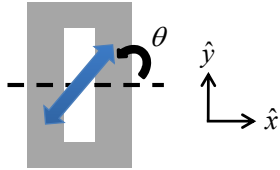  (b) 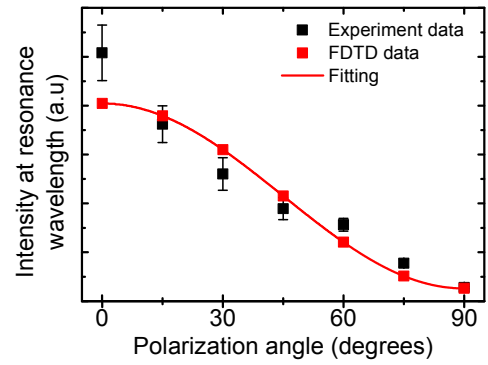

**Supplementary Figure S1.** (a) The polarization angle is taken as the angle between the electric field and the transverse direction of a slot. (b) Measured and calculated polarization-dependence of the transmission intensity at the resonance wavelength of a Ag single slot. The intensity follows $I(\theta) = I_0 \cos^2(\theta)$ (red line).

(a) 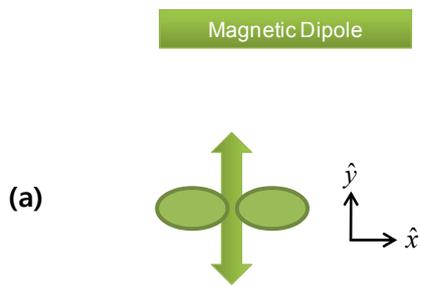

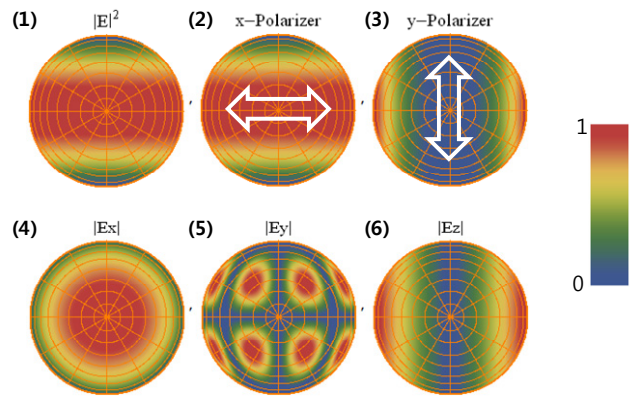

(b) 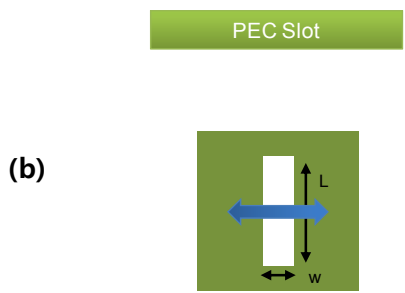

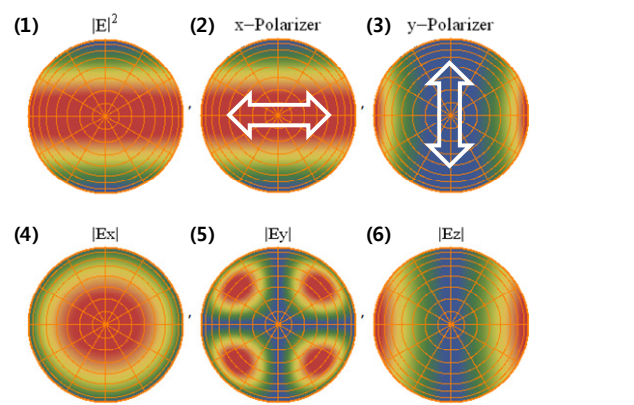

(c) 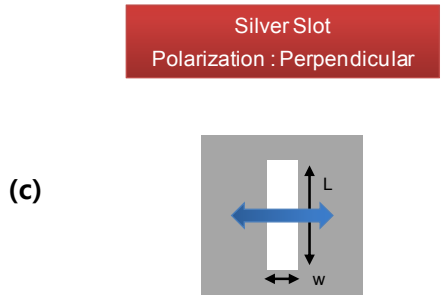

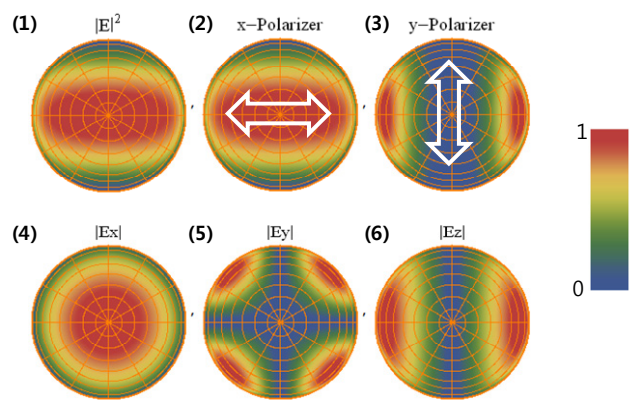

(d) 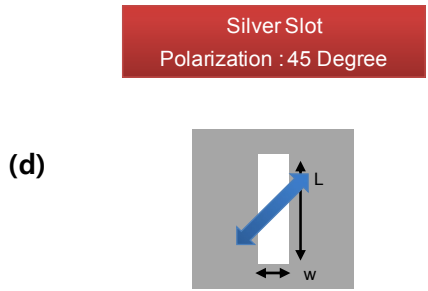

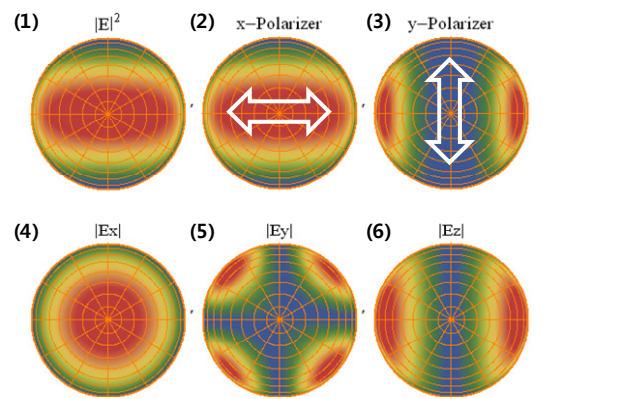

**Supplementary Figure S2.** Simulated far-field radiation patterns of (a) an ideal magnetic dipole, (b) a perfect electric conductor (PEC) rectangular slot irradiated by a perpendicularly polarized plane wave, (c) a Ag rectangular slot excited by a perpendicularly polarized plane wave, and (d) a Ag rectangular slot excited by a 45°-rotated polarized plane wave. As seen from (c)-(2) and (d)-(2), the rectangular Ag slot can be considered as a single magnetic dipole even when the slot is irradiated by a 45° polarized plane wave. To show the field component and intensity polarization effect, we plot (1) $|\vec{E}|^2$, (2) $|\vec{E}|^2$ with an *x*-polarizer, (3) $|\vec{E}|^2$ with a *y*-polarizer, (4) $|E_x|$, (5) $|E_y|$, and (6) $|E_z|$ captured at a 1-m distance from the radiators. The white arrows indicate the polarization of the detectors, and each plot is normalized. The amplitude of the far-field electric component is $|E_x| \geq |E_z| \gg |E_y|$. Therefore, the contribution of the *y*-component of the electric field to the total intensity is minimal.

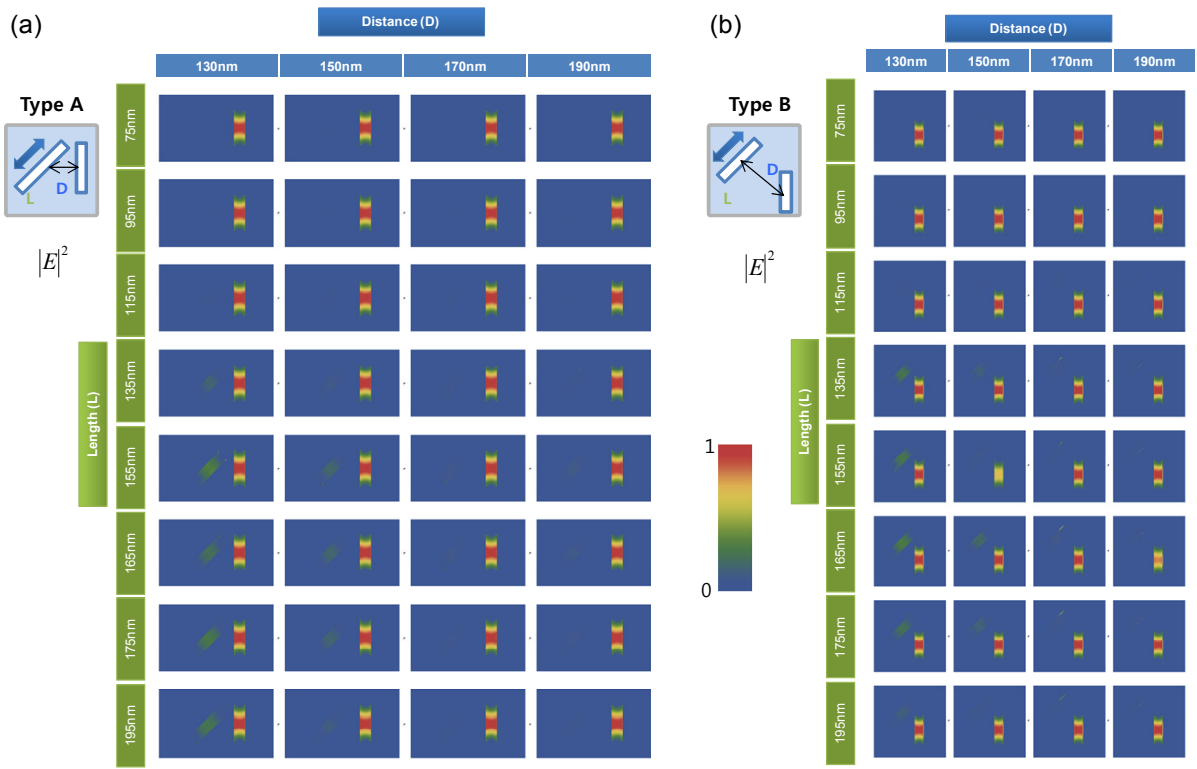

**Supplementary Figure S3.** Map of the normalized modal intensity profiles of two-slot antennas of types (a) A and (b) B. Intensity profiles for different L (D) values are vertically (horizontally) arranged. When the slots are irradiated by a polarized plane wave at 664 nm (denoted by a tilted blue double-headed arrow), fundamental electromagnetic modes are formed inside the feed slot regardless of the adjacent slot for both antenna types.

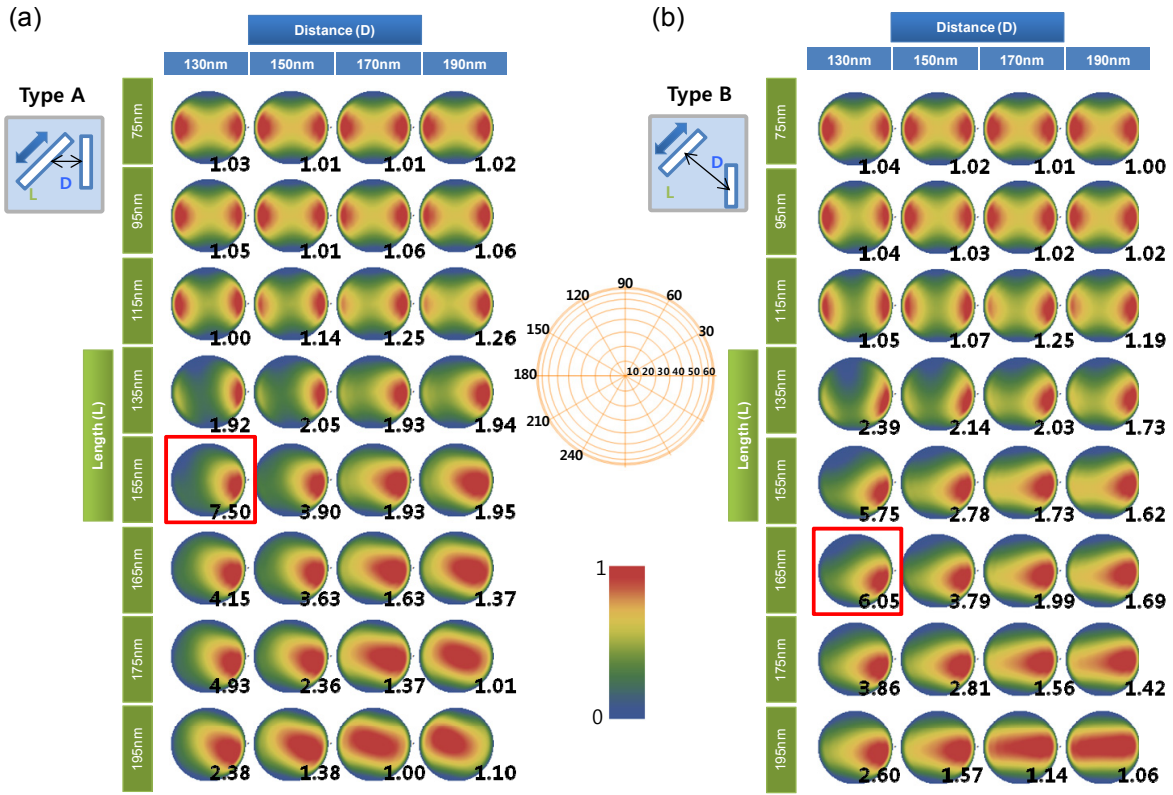

**Supplementary Figure S4.** Map of the normalized far-field intensities in the upper hemisphere with a 1-m radius for two-slot antennas of types (a) A and (b) B for various D and L values. The numbers in each plot are FB ratios. An optimal reflector for type A can be obtained with L = 155 nm and D = 130 nm (red square box in (a)) and that for type B with L = 165 nm and D = 130 nm (red square box in (b)). However, it is difficult to find good director conditions for the two-slot case. We can find a "weak" director condition with L = 95 nm and D = 130 nm for both antenna types with small FB ratios close to unity.

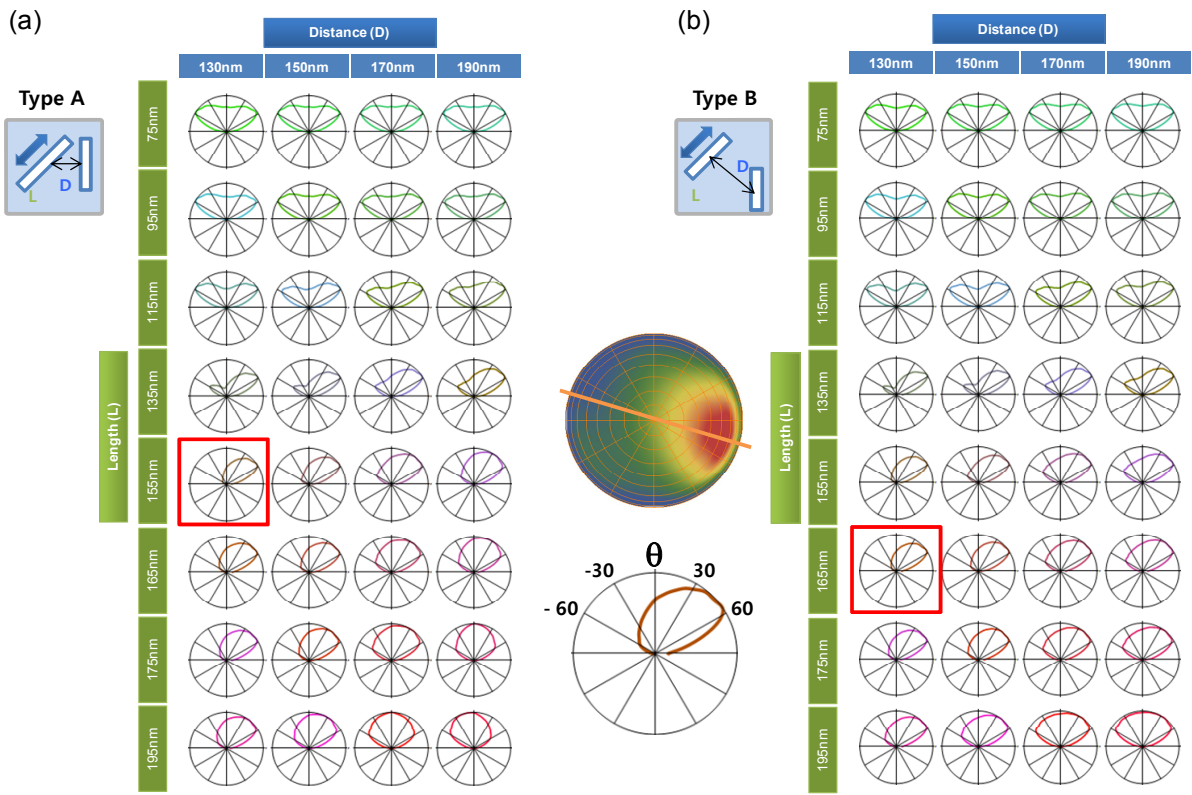

**Supplementary Figure S5.** At the maximum lobe intensity, we draw $\theta$-dependent plots for antenna types (a) A and (b) B by taking profiles of the normalized far-field density plots along the orange in the middle figure. Red square boxes represent the best reflector conditions for types A and B.

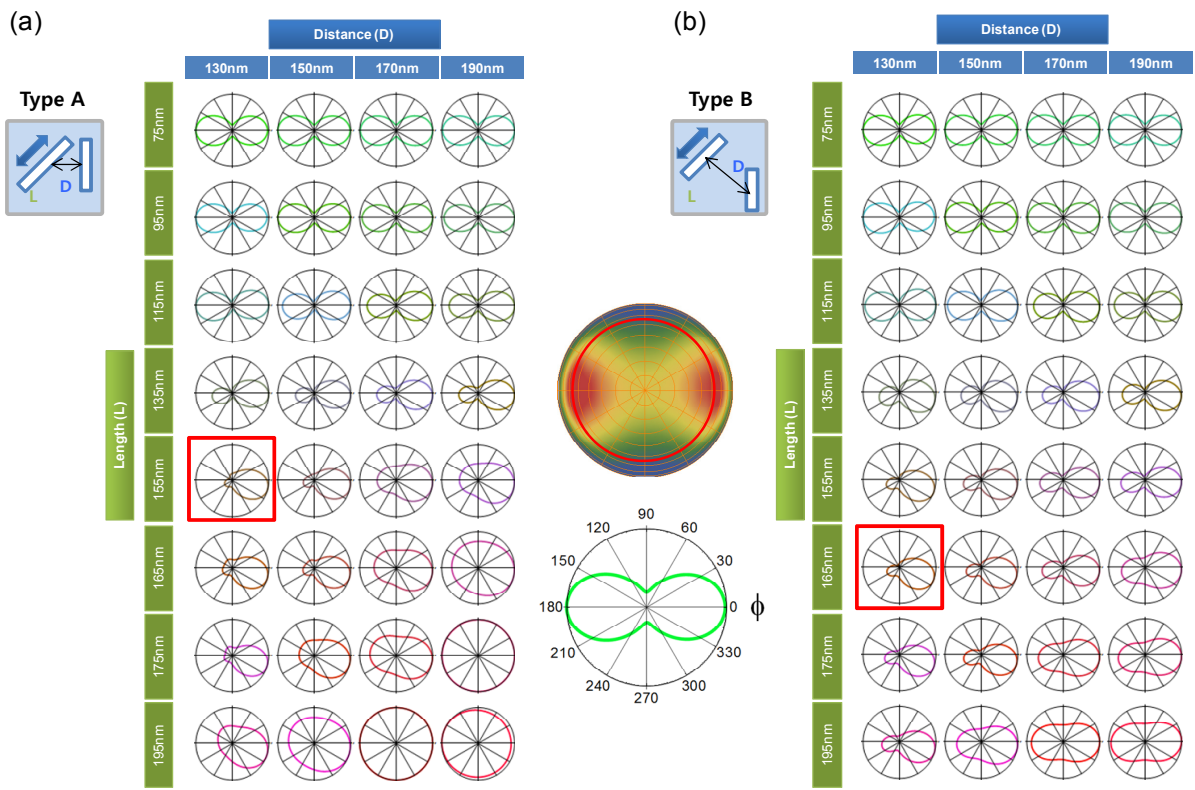

**Supplementary Figure S6.** At the maximum lobe intensity, we draw $\phi$-dependent plots for antenna types (a) A and (b) B by taking profiles of the normalized far-field density plots along the red line in the middle figure. Red square boxes represent the best reflector conditions for types A and B.

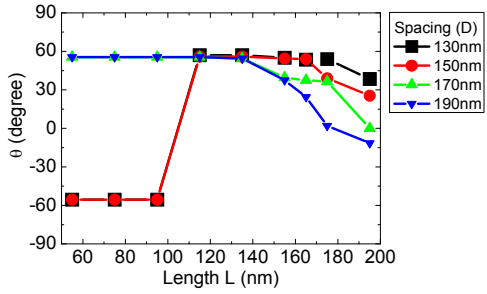 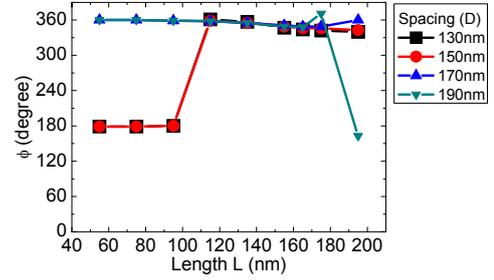

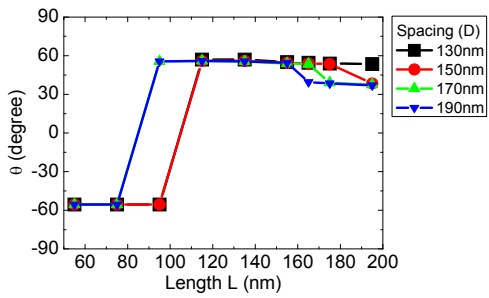 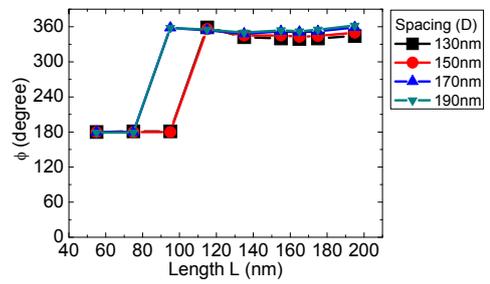

**Supplementary Figure S7.** The (a) θ- and (b) ϕ-coordinates at which the maximum lobe intensity occurs in each type-A antenna as a function of L and D. (c) and (d) Corresponding θ- and ϕ-coordinates for each type-B antenna.

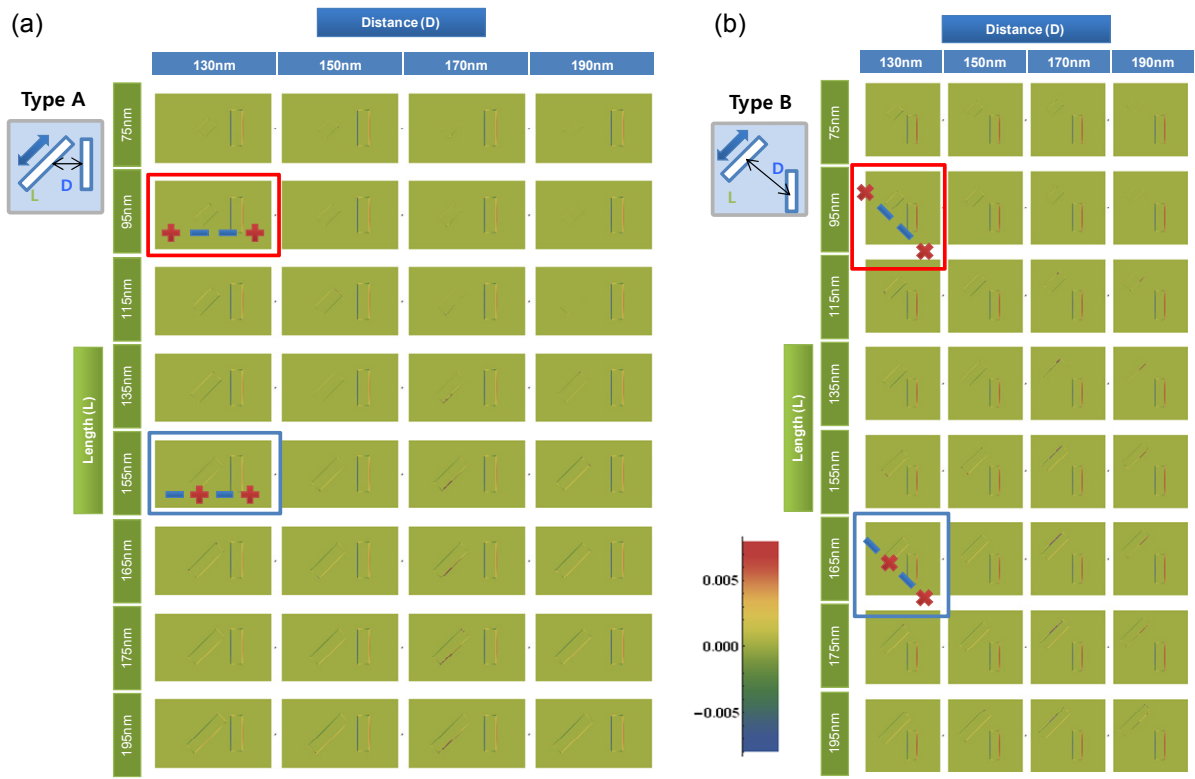

**Supplementary Figure S8.** Map of the charge distribution for antenna types (a) A and (b) B. The best reflector and director configurations have capacitive and inductive charge couplings between the two slots, respectively. The red (blue) box corresponds to the best director (reflector). This coupling feature is exactly opposite to the case of a multiple-rod-based antenna, as expected from Babinet's principle.

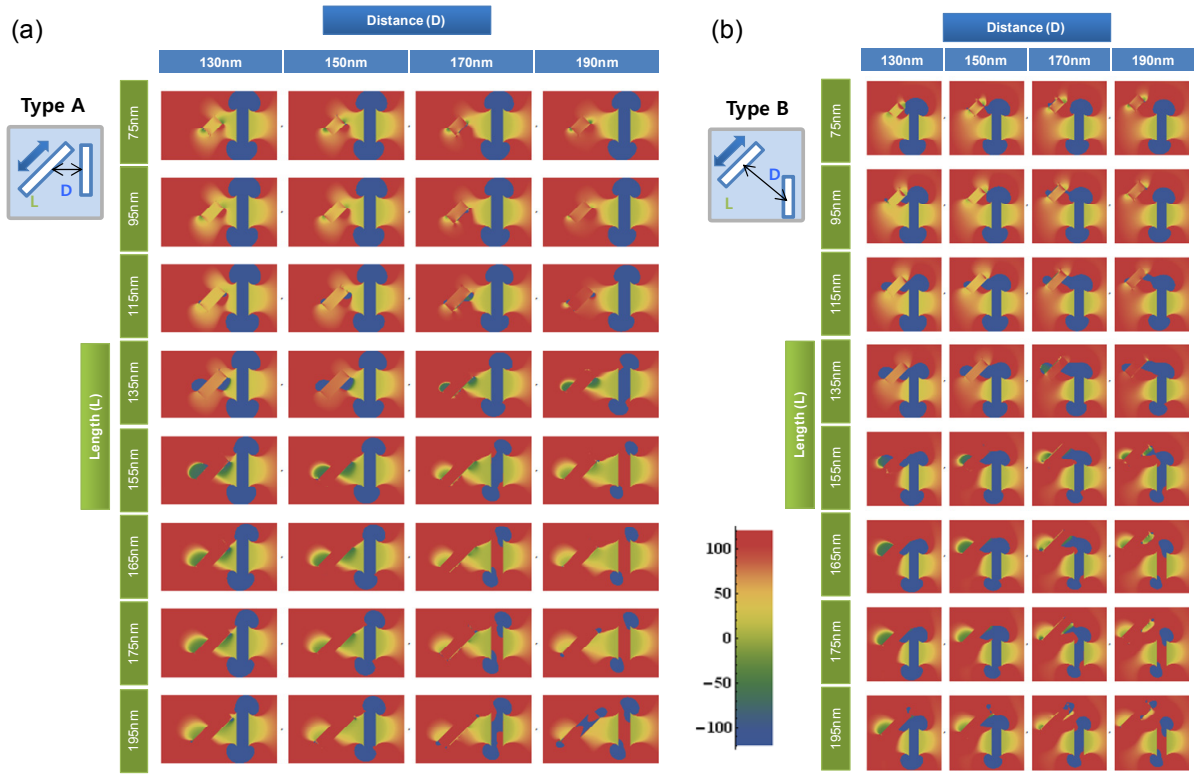

**Supplementary Figure S9.** Map of the phase distribution of the *x*-component of the electric field. The phase is almost uniform inside a slot, which enables us to define a single-valued phase measured at the centre of each slot. Furthermore, each slot can be regarded as a single magnetic dipole object even with an interaction between slots.

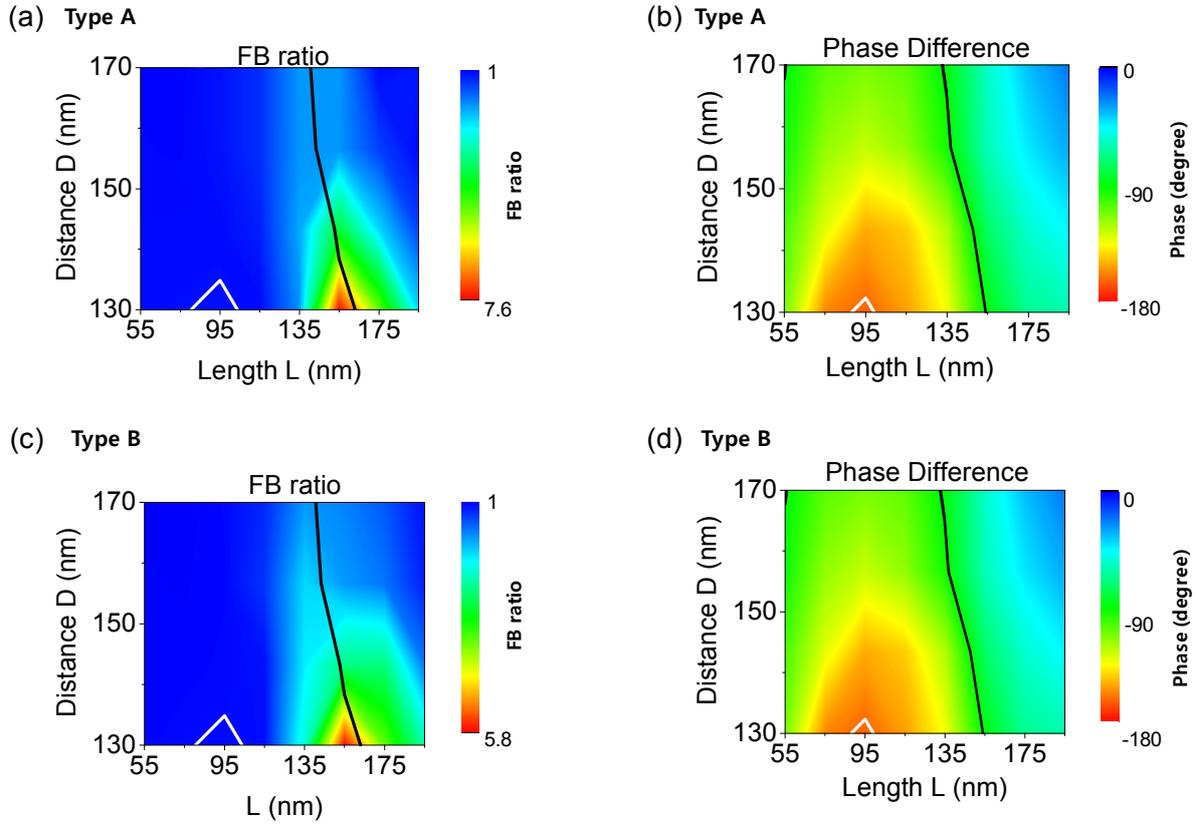

**Supplementary Figure S10.** Contour plots of the FB ratios and phase differences for antenna types (a), (b) A and (c), (d) B, respectively. We found that the phase difference, which is defined as the phase of the 2nd slot element minus that of the feed, and the FB ratio have a close relationship. We observe that the optimal reflector condition (D = 130 nm and L = 155 nm for type A and D = 130 nm and L = 165 nm for type B) overlaps with the −90° phase difference line (black lines). For the director, the best phase condition (D = 130 nm and L = 95 nm for both types A and B) allowable in our fabrication range is approximately −170° (white lines), which is insufficient for the optimal director phase of −270°.

(a) **Type A** 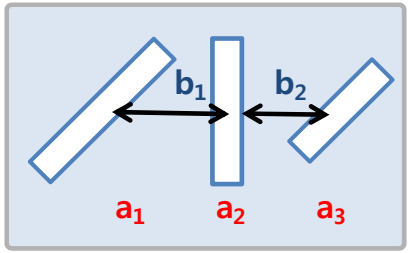

(b) **Type B** 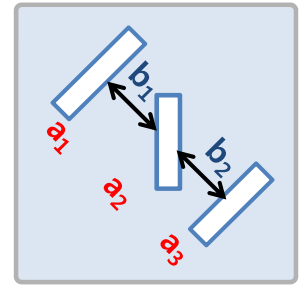

(c) 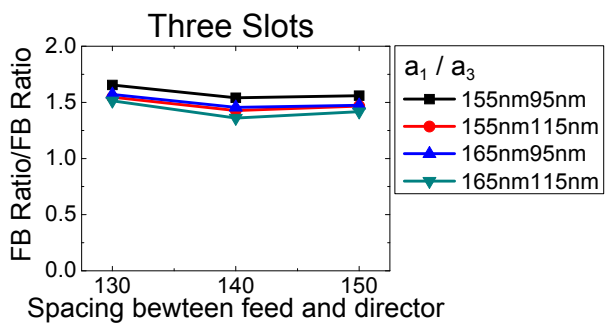

**Supplementary Figure S11.** Various parameters of the (a) type-A and (b) type-B three-slot nanoantennas. (c) The ratios of the simulated FB ratios (type A/type B) for the three-slot antennas as a function of $b_2$ with fixed $b_1$ = 130 nm and $a_2$ = 145 nm. In all cases, type A provides better FB ratios than type B.

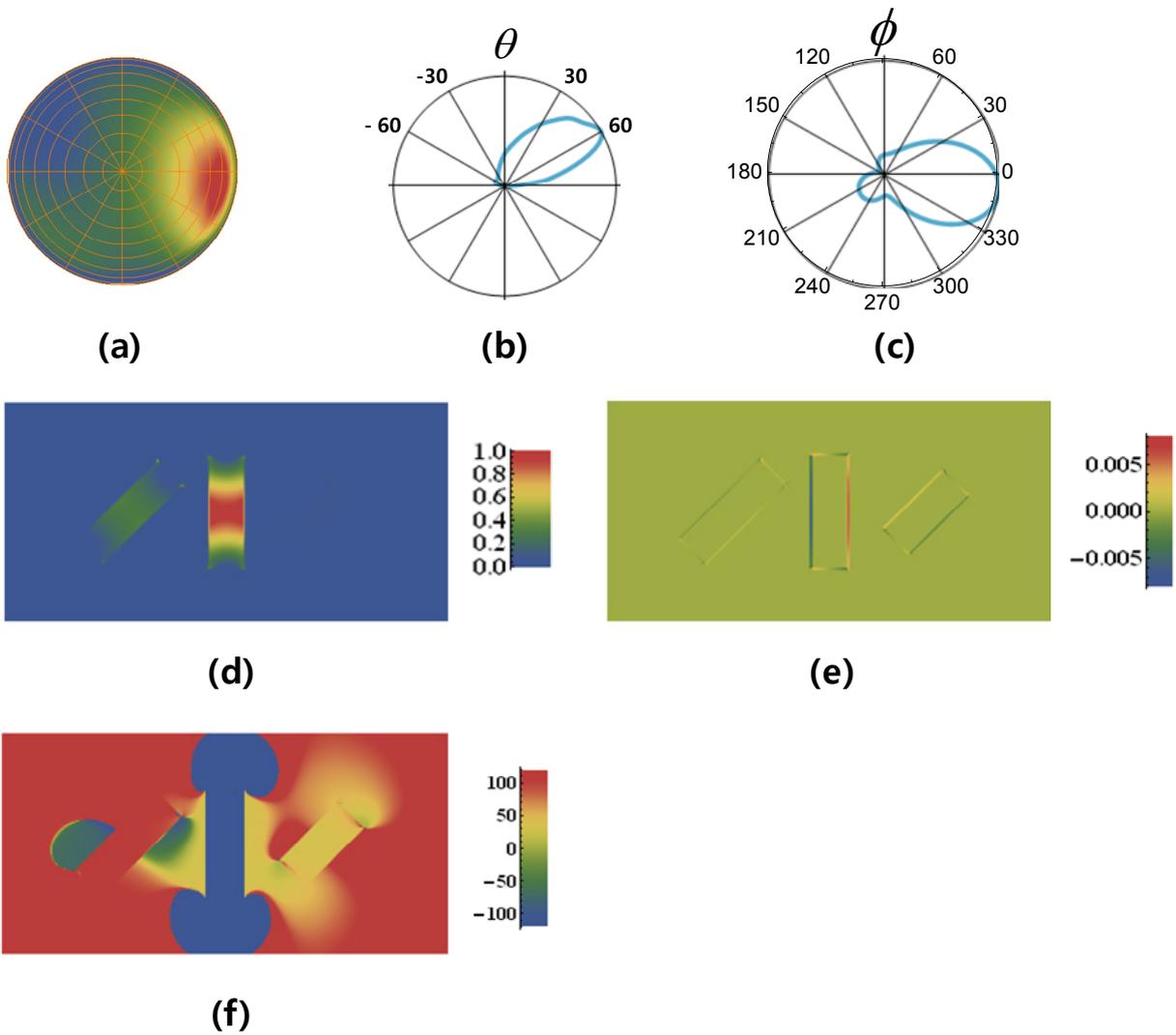

**Supplementary Figure S12.** (a) Density plot of the radiation pattern of a type A antenna when $a_1 = 165$ nm, $a_3 = 95$ nm, and $b_2 = 140$ nm. The corresponding (b) $\theta$ and (c) $\phi$ plots. Distributions of the (d) electric field intensity, (e) charge, and (f) phase.

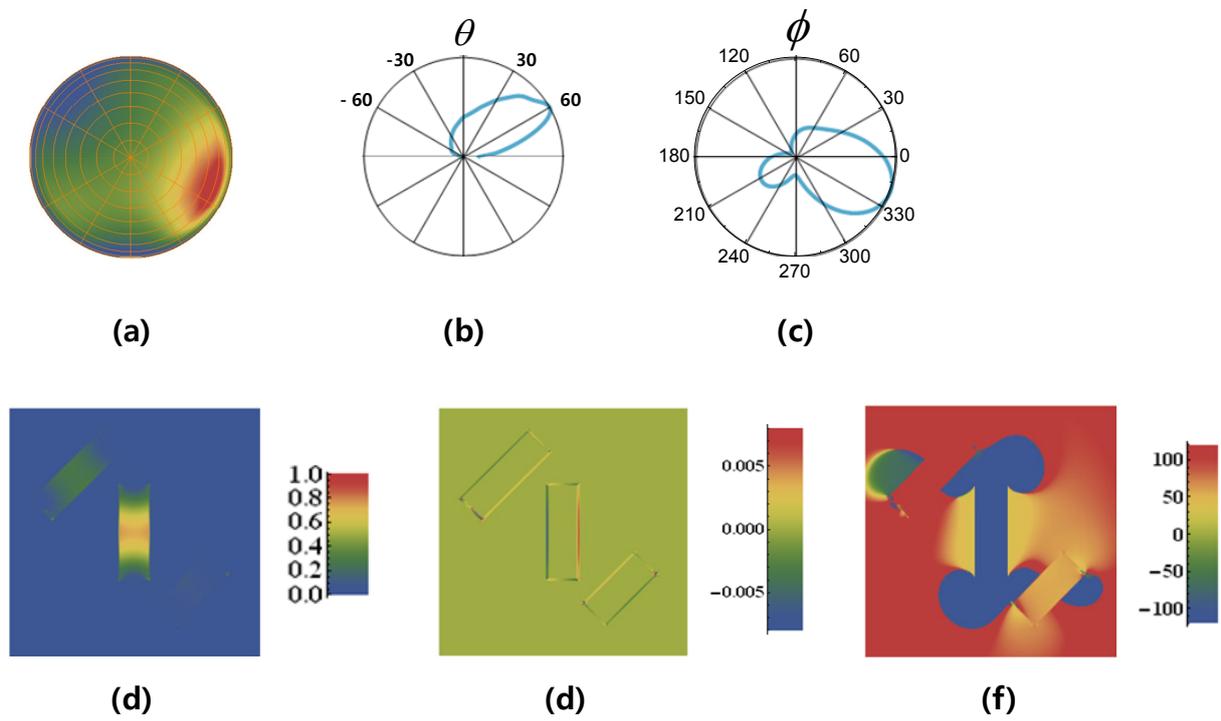

**Supplementary Figure S13.** (a) Density plot of the radiation pattern of a type B antenna when $a_1$ = 165 nm, $a_3$ = 95 nm, and $b_2$ = 150 nm. The corresponding (b) θ and (c) ϕ plots. Distributions of the (d) electric field intensity, (e) charge, and (f) phase.

(a) Type A

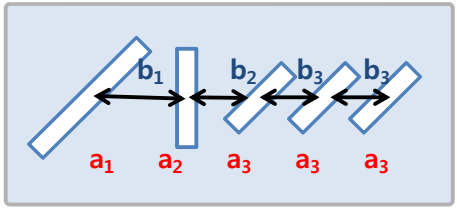

(b) Type B

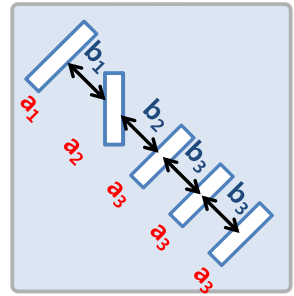

(c)

| | $b_1$ (nm) | $b_2$ (nm) | $b_3$ (nm) |
|---|---|---|---|
| 1 | 130 | 130 | 130 |
| 2 | 130 | 130 | 150 |
| 3 | 130 | 150 | 150 |
| 4 | 130 | 150 | 170 |
| 5 | 140 | 140 | 140 |
| 6 | 140 | 140 | 160 |
| 7 | 140 | 160 | 160 |
| 8 | 140 | 160 | 180 |

(d)

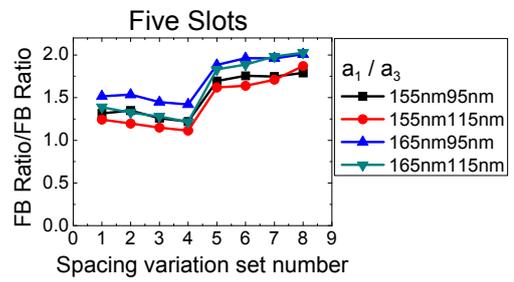

**Supplementary Figure S14.** Various parameters of the (a) type-A and (b) type-B five-slot nanoantennas. (c) Table of the distances between the elements chosen for the simulation. (d) The ratios of the simulated FB ratios (type A/type B) for the five-slot antennas for a given feed length of $a_2$ = 145 nm. In all cases, type A provides better FB ratios than type B.

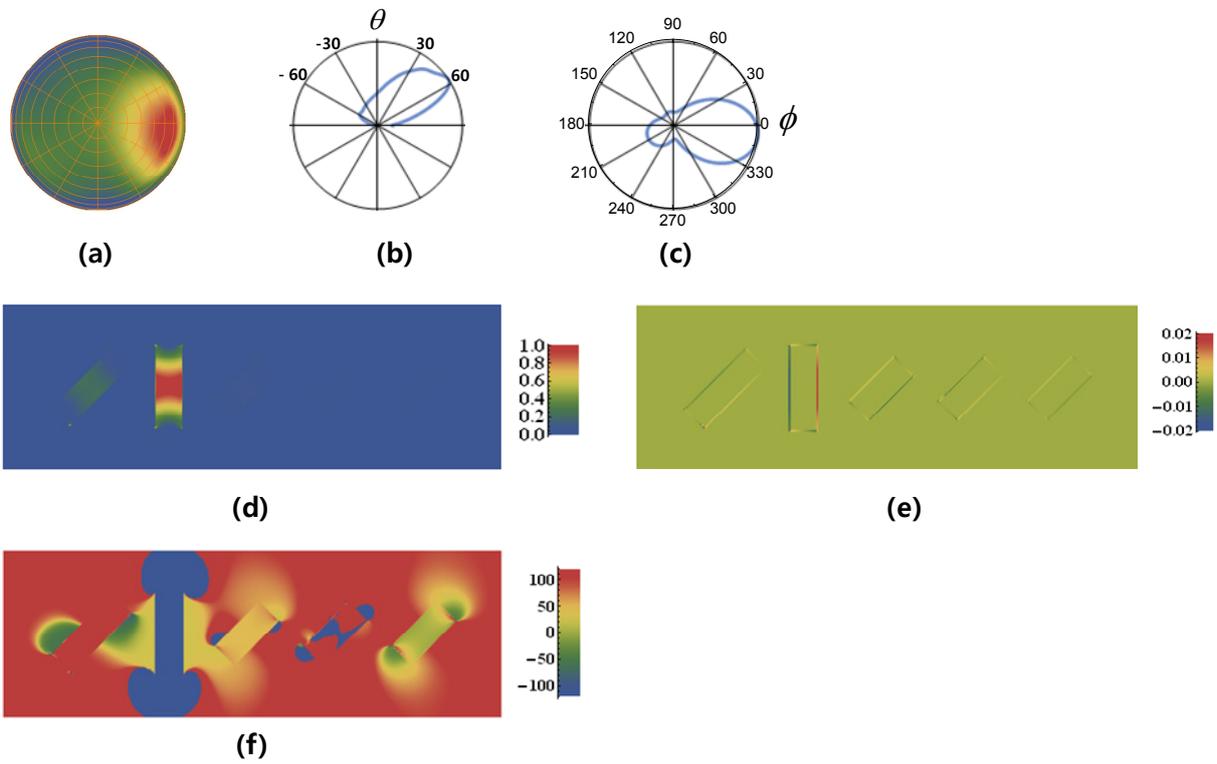

**Supplementary Figure S15.** (a) Density plot of the radiation pattern of a type-A five-slot antenna. The corresponding (b) θ and (c) ϕ plots. Distributions of the (d) electric field intensity, (e) charge, and (f) phase.

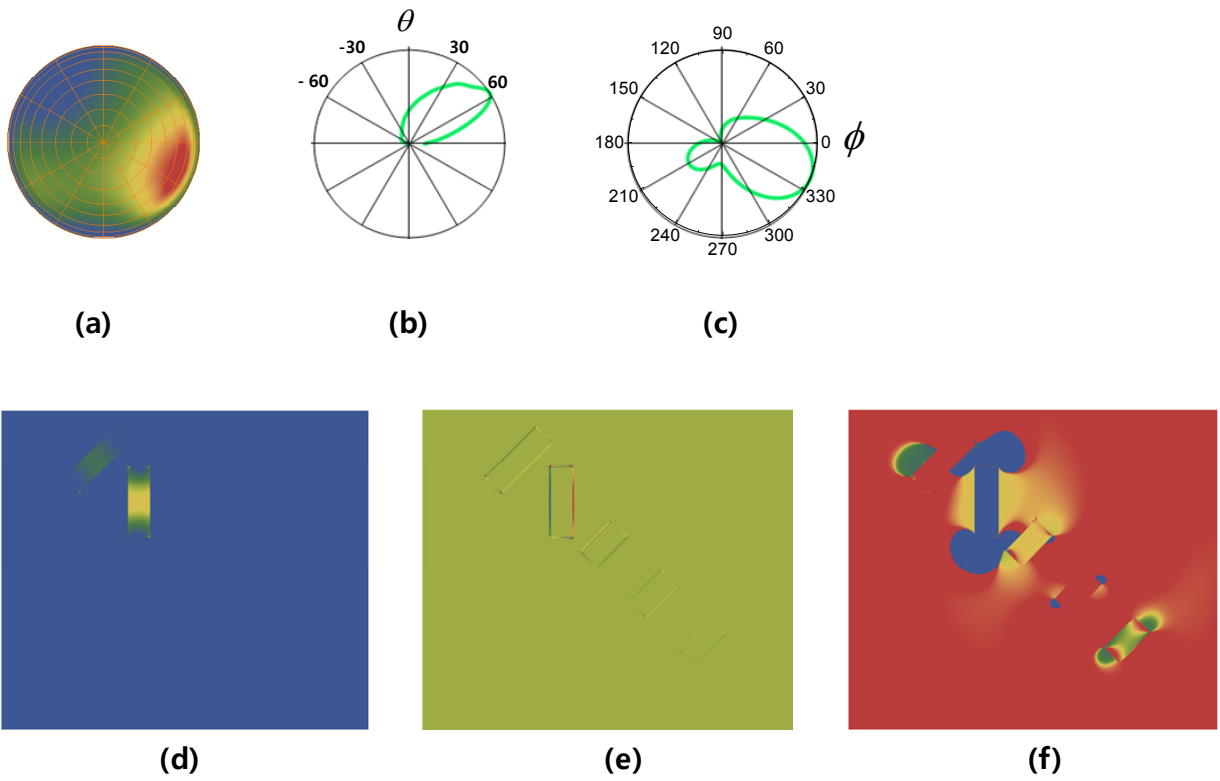

**Supplementary Figure S16.** (a) Density plot of the radiation pattern of a type-B five-slot antenna. The corresponding (b) θ and (c) ϕ plots. Distributions of the (d) electric field intensity, (e) charge, and (f) phase.

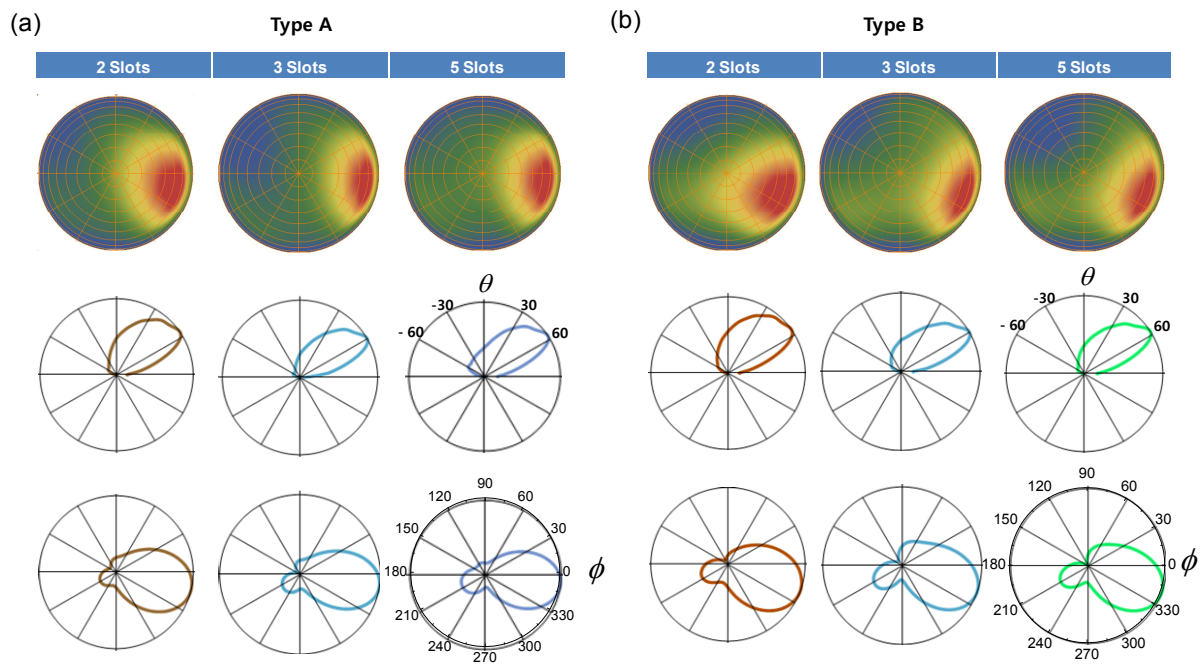

**Supplementary Figure S17.** Comparison between the two-, three-, and five-slot antennas of types A and B. (a) Radiation patterns (first row), θ plot (second row), and ϕ plot (third row) for type-A antennas. With an increasing number of slots, the radiation pattern becomes narrower with an increasing FB ratio. (b) Radiation patterns (first row), θ plot (second row), and ϕ plot (third row) for type-B antennas.